\documentclass[preprint,12pt,a4paper]{elsarticle}
\journal{European Journal of Mechanics B/Fluids}
\usepackage[latin1]{inputenc}
\usepackage{amsmath,amssymb}
\usepackage{graphicx}
\renewcommand\d{\mathrm{d}}
\renewcommand\Re{\mathrm{Re}}
\DeclareMathOperator{\e}{e}
\newcommand\be{\begin{equation}}
\newcommand\ee[1]{\label{#1}\end{equation}}
\newcommand\ie{\textit{i.e.}}
\newcommand\eg{\textit{e.g.}}
\newcommand\etal{\textit{et al.}}

\title{Structure and interpolation of the turbulent velocity profile in parallel flow}
\author{Paolo Luchini}
\address{Universit\`a di Salerno, DIIN, %
84084 Fisciano (SA), Italy}
\ead{luchini@unisa.it}
\begin{document}
\begin{abstract}
The classical scaling theory of turbulent parallel flow provides a framework for the description of the mean velocity profile through two functions of one variable, traditionally named law of the wall and law of the wake, and a universal logarithmic law characterized by von Kármán's constant. Despite its widespread adoption in research and in teaching, discrepancies between this theory and both experiments and numerical simulations have been repeatedly observed in the literature. Recently we have shown that in the logarithmic layer such discrepancies can be physically interpreted and analytically accounted for through an equally universal correction caused by the pressure gradient. This finding opens the way to a likewise improvement in the description of the law of the wall and of the law of the wake, an analytical interpolation of either of which is often useful for practical applications.
\end{abstract}
\maketitle

\section{Background and introduction}
The classical scaling theory of the turbulent velocity profile, based after Prandtl, von K\'arm\'an and Millikan on the distinction between a wall layer and a defect layer, is a mainstay of turbulence theory. It underlies the friction law of pipes and channels and also constitutes a basic ingredient of the description of the turbulent boundary layer. Therefore, the extraction of the empirical coefficients (and functions, such as the universal law of the wall and Coles' law of the wake) contained in this theory from experimental data and numerical simulations is a continuing activity that absorbs the attention of scientists. At some point, however, the questions must arise (and have repeatedly arisen in the literature) of the level of precision with which the Prandtl-vonK\'arm\'an-Millikan theory (which of course was originally conceived as, and cannot be other than, an approximation) is valid, how quickly it is approached in the limit of Reynolds number tending to infinity, and whether it can be improved with higher-order corrections in some kind of formal asymptotic expansion. It was recently observed \cite{PRL} that a correction proportional to the pressure gradient, considered as a small perturbation, can explain the difference between circular-pipe, plane-parallel-duct and plane-Couette flow and accord them to a single universal law. Purpose of the present article is to explore how much more can be extracted, from numerical and experimental data available today, in the way of an answer to the other general questions, and how much of this answer can be captured in an analytical interpolation of empirical data.

We must preliminarily remark that an analysis purporting to extract finer and finer corrections purely from empirical data must look at the data at increasing levels of magnification, and will soon meet a point where what is seen is no longer data but just either systematic or random error. One must therefore resist the temptation to overinterpret the data.
Doing so necessarily involves subjective judgement, and will force us to make some bold conjectures about what in the available data constitutes the physical phenomenon and what constitutes error. Like all conjectures, some of these may in the future turn out to be wrong, and what is identified here as error may eventually turn out to be an even finer physical feature (or viceversa). I am sure the authors of those data will understand that the term ``error"
is not intended to convey any negative connotation and is used here in the same sense as in the theory of measurement error.

Nevertheless, as will be clear at the end of this paper, the velocity profiles extracted from today's available numerical simulations and experiments are affected by a quantifiable error which unfortunately is insufficiently small for some parts of the present analysis. While this detracts nothing from the considerable effort and patience put by the authors in their computations and experiments (an error bar must well be placed somewhere), it does point out the usefulness to pursue further refinements. 

\begin{table}
\caption{Curve labelling and data sources}
\label{plotdata}
\vspace{1em}
\centering
\begin{tabular}{lclcc}
Acronym & Geometry & $\Re_\tau$ & Technique & Reference\\
\hline
HJ550 & plane & 547 & DNS & \cite{Jimenez}\\
HJ1E3 & plane & 934 & DNS & \cite{Jimenez}\\
HJ2E3 & plane & 2004 & DNS & \cite{Jimenez}\\
DJ4E3 & plane & 4179 & DNS & \cite{Jimenez4000}\\
LM550 & plane & 543 & DNS & \cite{Moser}\\
LM1E3 & plane & 1001 & DNS & \cite{Moser}\\
LM2E3 & plane & 1995 & DNS & \cite{Moser}\\
LM5E3 & plane & 5186 & DNS & \cite{Moser}\\
BPO550 & plane & 550 & DNS & \cite{Bernardini}\\
BPO1E3 & plane & 999 & DNS & \cite{Bernardini}\\
BPO2E3 & plane & 2021 & DNS & \cite{Bernardini}\\
BPO4E3 & plane & 4079 & DNS & \cite{Bernardini}\\
Kea360 & pipe & 361 & DNS & \cite{Schlatter}\\
Kea550 & pipe & 550 & DNS & \cite{Schlatter}\\
Kea1E3 & pipe & 999 & DNS & \cite{Schlatter}\\
WM700 & pipe & 685 & DNS & \cite{Moin}\\
WM1E3 & pipe & 1142 & DNS & \cite{Moin}\\
N300 & pipe & 294 & Exp. & \cite{Nikuradse}\\
N700 & pipe & 657 & Exp. & \cite{Nikuradse}\\
N2500 & pipe & 2511 & Exp. & \cite{Nikuradse}\\
N4500 & pipe & 4537 & Exp. & \cite{Nikuradse}\\
N14000 & pipe & 14200 & Exp. & \cite{Nikuradse}\\
N35000 & pipe & 35200 & Exp. & \cite{Nikuradse}\\
N55000 & pipe & 55500 & Exp. & \cite{Nikuradse}\\
Sea2E3 & pipe & 1985 & Exp. & \cite{Smits}\\
Sea3E3 & pipe & 3334 & Exp. & \cite{Smits}\\
Sea5E3 & pipe & 5411 & Exp. & \cite{Smits}\\
Sea1E4 & pipe & 10480 & Exp. & \cite{Smits}\\
Sea2E4 & pipe & 20250 & Exp. & \cite{Smits}\\
Sea4E4 & pipe & 37690 & Exp. & \cite{Smits}\\
Sea7E4 & pipe & 68160 & Exp. & \cite{Smits}\\
Sea1E5 & pipe & 98190 & Exp. & \cite{Smits}\\
PBO260 & couette & 260 & DNS & \cite{Pirozzoli}\\
PBO500 & couette & 507 & DNS & \cite{Pirozzoli}\\
PBO1E3 & couette & 986 & DNS & \cite{Pirozzoli}\\
\end{tabular}
\end{table}
In this paper extensive use will be made of velocity profiles taken from the literature, which will be tagged by authors' initials and Reynolds number. Tags and the references to be credited for each profile are collected in Table \ref{plotdata}. We warmly thank all the authors involved for making their precious data files publicly available on the web.

\subsection{The classical theory of parallel turbulent flow}
The general time-averaged, fully developed velocity profile $u(z)$ in an infinitely long straight duct of a given cross section is a function of the pressure gradient $p_x$, density $\rho$, kinematic viscosity $\nu$ and a characteristic height $h$ of the cross-section. Momentum balance links the pressure gradient to the mean wall shear stress $\tau_w$ as
\be
-p_x=4\tau_w/D_H
\ee{momentum}
where $D_H$, the hydraulic diameter, is 4 times the ratio of area to perimeter of the cross section. It may sometimes be convenient to assume $D_H$ itself as the characteristic dimension $h$, but for the time being we shall let them be distinct. $h$ will be chosen to be the distance from the wall to the symmetry axis, \ie\ half the distance between walls of a plane duct or the radius of a pipe.
Nondimensionalization allows the velocity profile to be expressed as a function of only two variables:
\be
u^+=u^+(z^+;h^+)
\ee{genprofile}
where $u^+=u/u_\tau$, $u_\tau=\sqrt{\tau_w/\rho}$, $z^+=zu_\tau/\nu$, and $h^+= hu_\tau/\nu$, quantities denoted by a $^+$ being also commonly denoted as measured in ``wall units". The dimensionless height $h^+$ in wall units coincides with the shear-based Reynolds number $\Re_\tau= hu_\tau/\nu$; one or the other name will be more explicative depending on context.

The classical asymptotic theory of the turbulent velocity profile \cite[see also 13]{Millikan} distinguishes three regions in the limit of $\Re_\tau=h^+\rightarrow\infty$:
\begin{itemize}
\item a ``wall'' layer, for $z^+\ll h^+$, where $u^+$ is a function of $z^+$ only,
\be 
u^+\simeq f(z^+);
\ee{claswall}
\item a ``defect" layer, for $z^+\gg 1$, where the velocity defect $u^+(h^+)-u^+(z^+)$ is a function of $Z=z/h=z^+/h^+$ only,
\be
u^+\simeq u^+(h^+)-F(Z);
\ee{clasdef}
\item an ``overlap" layer, for $1\ll z^+ \ll h^+$, where both \eqref{claswall} and \eqref{clasdef} must simultaneously be valid. According to the standard argument by Millikan \cite{Millikan}, their simultaneous validity implies that in the overlap layer the velocity profile is logarithmic,
\be
u^+\simeq A \log(z^+)+B = A\log(Z)+A\log(h^+)+B,
\ee{clasovlap}
where the reciprocal of the $A$ coefficient is known as von K\'arm\'an's constant $\kappa=1/A$.
\end{itemize}
In addition, equations \eqref{claswall} and \eqref{clasovlap}, operating in a layer much thinner that the characteristic height $h^+$, must also be independent of geometry, which is why \eqref{claswall} is named the ``universal law of the wall". By contrast function $F(Z)$ in \eqref{clasdef} is allowed to vary from one geometry to another.

The layered structure of the turbulent velocity profile can also be recast as a uniformly valid approximation, by summing \eqref{claswall} and \eqref{clasdef} and subtracting their common asymptotic behaviour \eqref{clasovlap}, \ie
\[
u^+ = f(z^+) +u(h^+) - F(Z) -A\log(Z)-A\log(h^+)-B,
\]
or equivalently

\be
u^+=f(z^+)+G(Z).
\ee{clasunif}
$G(Z)$ represents the deviation of velocity in the defect layer from the logarithmic law, defined as
\be
 G(Z)=-F(Z)-A \log(Z)+C,
\ee{Gdef}
with $C=\lim_{Z\rightarrow 0}[F(Z)+A \log(Z)]$.
By definition of $C$ (not the same $C$ as used in \cite{PRL}), one has $G(0)=0$. It also follows that $G(1)=C$ [because from \eqref{clasdef} $F(1)=0$] and that
\be
u^+(h^+)=\kappa^{-1} \log(h^+)+B+C.
\ee{clascntrvel}
This last relationship is at the foundation of the friction law linking friction coefficient to Reynolds number (\eg, \cite{Schlichting} p. 515).

Function $G(Z)$ is known as the ``law of the wake" from the classical work of Coles \cite{Coles}, who conjectured on an empirical basis that, in a boundary layer, $G(Z)$ might have a universal shape approximately similar to a mixing layer (or half of a wake), up to a multiplicative constant.

\subsection{Pressure-gradient correction to the overlap layer}
\label{pgcorr}
\begin{figure}
\includegraphics[width=\columnwidth]{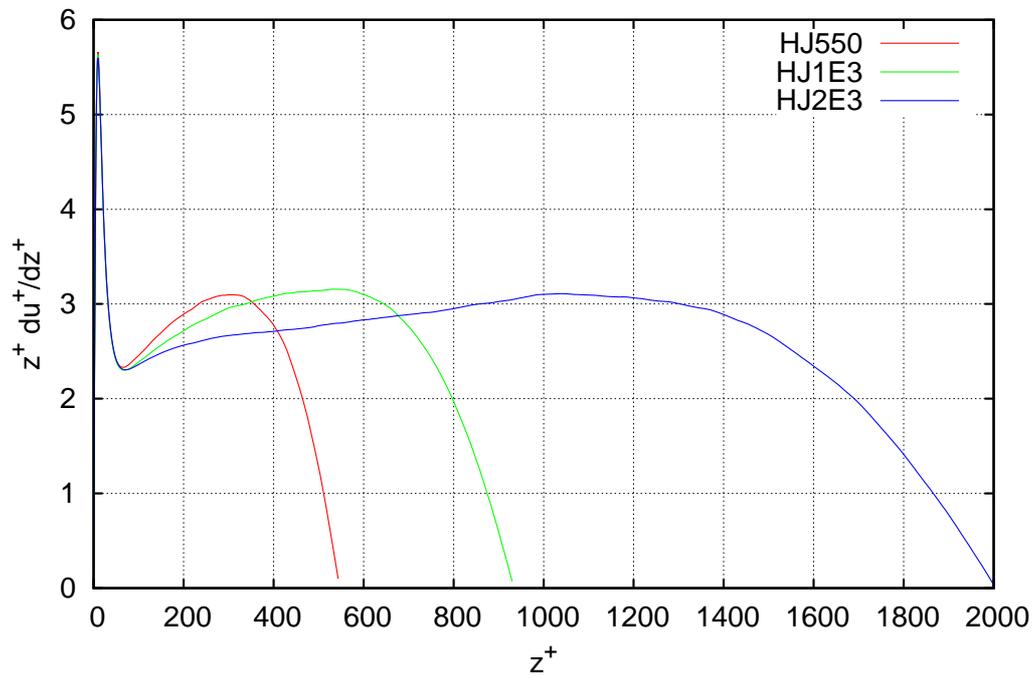}
\caption{logarithmic derivative of the velocity profile for DNS from the same source \cite{Jimenez} at different Reynolds numbers. According to the logarithmic law \eqref{clasovlap}, this function should approach a constant $A=\kappa^{-1}$ in some intermediate range to be identified with the overlap layer.}
\label{logderiv}
\end{figure}
Several authors have questioned the validity of the logarithmic law, and proposed alternatives, on the basis that \eqref{clasovlap} provides a somewhat unsatisfactory fit of empirical (both experimental and numerical) data; alternately one may also conclude that \eqref{clasovlap} is valid but the Reynolds number of present-day direct numerical simulations (or even experiments) is still insufficiently high to see the true asymptotic behaviour and estimate von K\'arm\'an's constant $\kappa$. Such pessimistic deductions are substantiated by plots of the logarithmic derivative of $u^+$, a sample of which is given in Figure \ref{logderiv}. According to the logarithmic law \eqref{clasovlap}, a plot of the derivative of $u^+$ against $\log z^+$ should flatten to a constant $A=\kappa^{-1}$ in some intermediate region to be identified with the overlap layer, and admittedly whether it does in Figure \ref{logderiv} can be doubtful.
\begin{figure}
\includegraphics[width=\columnwidth]{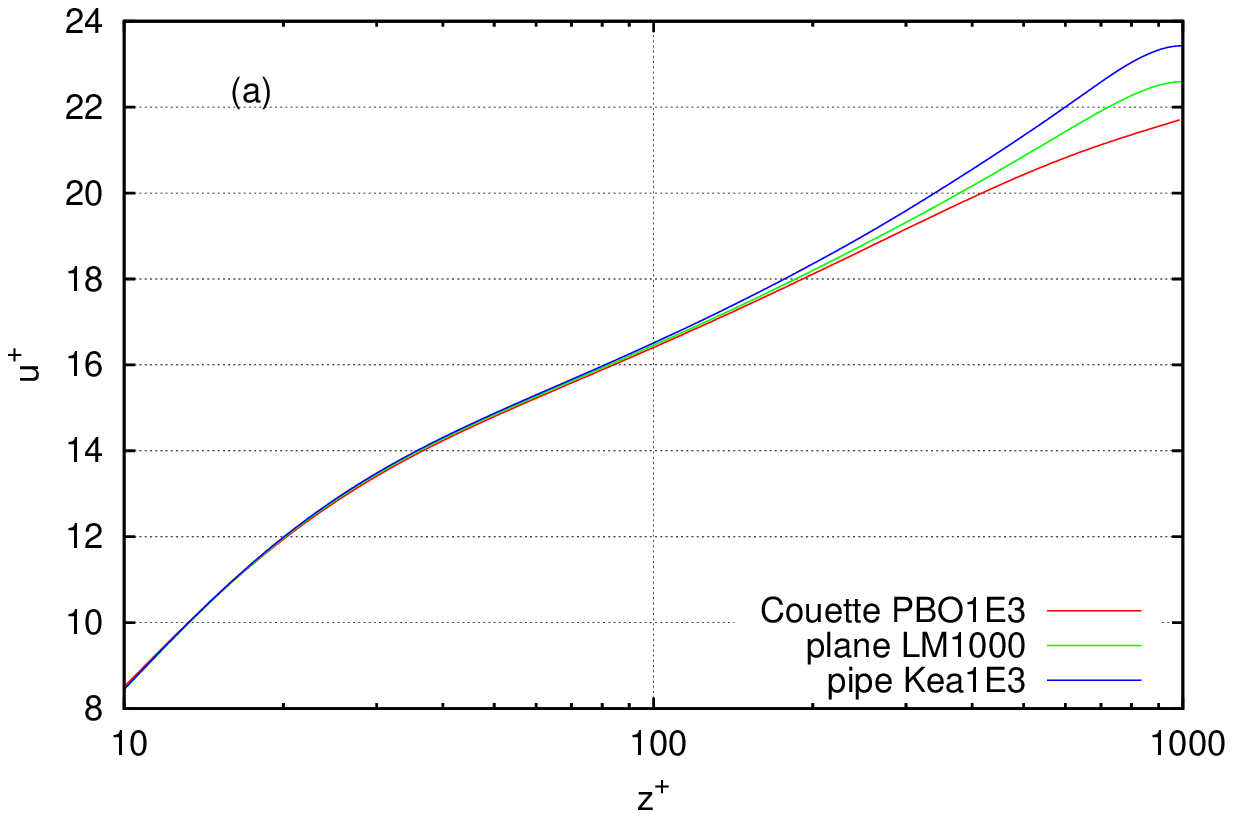}

\includegraphics[width=\columnwidth]{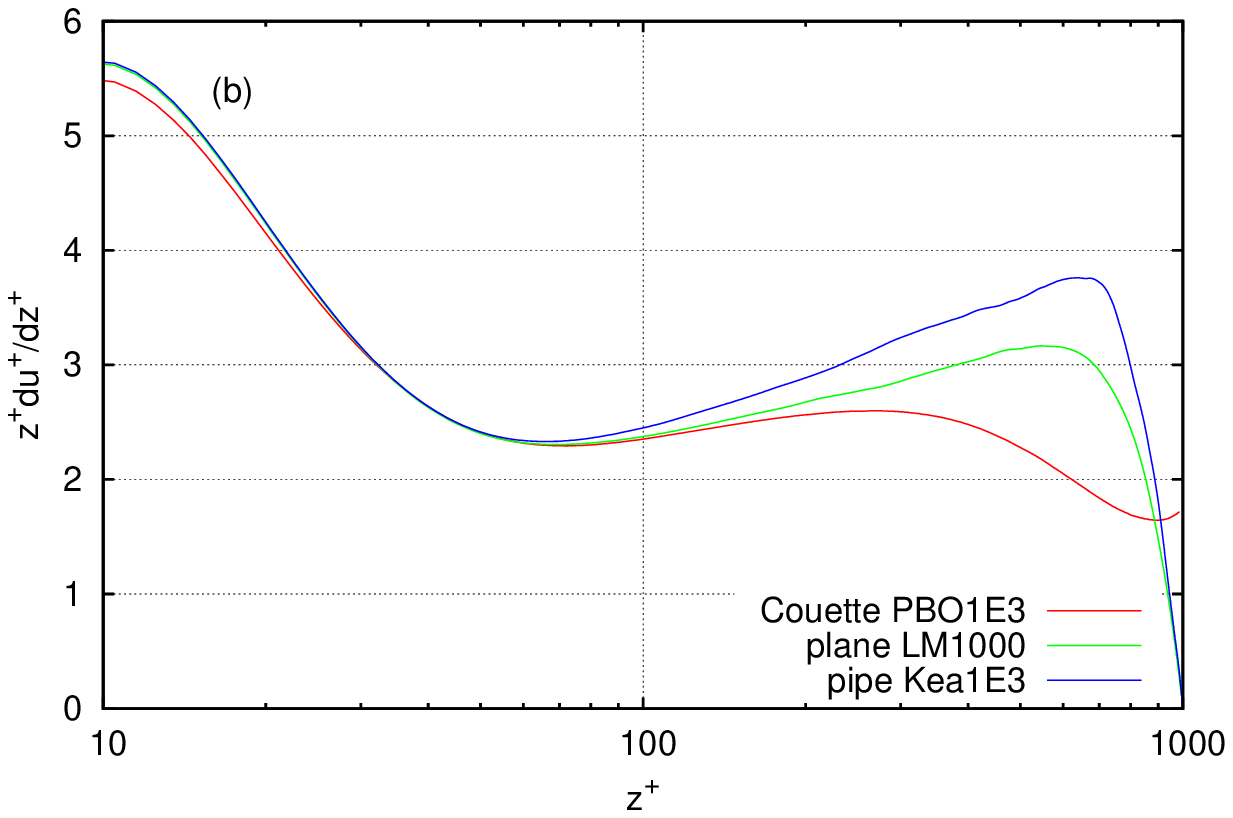}
\caption{Velocity profile in wall units versus wall-normal
coordinate on a logarithmic scale (a) and its logarithmic derivative (b), for three different geometries at $\Re_\tau=1000$.}
\label{threevel}
\end{figure}
More insight can be gained if, instead of plotting the velocity profile or its logarithmic derivative for the same geometry at different Reynolds numbers, we plot them at the same Reynolds number for different (circular or plane) geometries. This was done in Figure 1 of our previous article \cite{PRL}, reproduced as Figure \ref{threevel} here, plotting together three turbulent velocity profiles at the same Reynolds number $\Re_\tau=1000$, one for pipe flow from \cite{Schlatter}, one for plane duct flow from \cite{Moser} and one for plane Couette flow from \cite{Pirozzoli}. Not only the curves are not straight lines, as is even more clearly brought out in the companion plot of their logarithmic derivative, but in addition the maximum value of velocity is considerably different in each case and so is its average slope.
Upon looking at Figures \ref{logderiv}-\ref{threevel} one may:
\begin{enumerate}
\item conclude that the logarithmic law is invalid, or not yet valid for $\Re_\tau\le 1000$, except perhaps for Couette flow;
\item strike a straight line through the velocity plot, assuming that deviations will be somehow compensated, and choose an average slope value as the best approximation of the coefficient $\kappa^{-1}$, to be definite if very crude let us say
\be
\kappa^{-1}=[u^+(h^+)-u^+(50)]/[\log(h^+)-\log(50)];
\ee{kinter}
\item select the minimum at $z^+\simeq 70$ common to Figures \ref{logderiv} and \ref{threevel}b, which is also the largest value of $z^+$ where all the curves more or less coincide, as the ``true" value of $\kappa^{-1}$ and assume that this minimum will keep the same value as the profile becomes flatter and flatter with increasing $\Re_\tau$ (a trend made plausible by a comparison of velocity profiles with changing $\Re_\tau$ such as in Figure \ref{logderiv}).
\end{enumerate}

Route 1 is probably the one that over time elicited the strongest opinions. %
Considerations about the lack of straightness of logarithmic velocity profiles similar to Figures \ref{logderiv} and \ref{threevel} have emerged from both experimental data (extending to much larger Reynolds numbers) and numerical simulations. Alternatives to the logarithmic law have been investigated, the most influential and articulate proponent probably being Barenblatt \cite{Barenblatt}.

Route 2 leads to the conclusion (championed by \cite{Chauhan,Nagib}, although with a subtler argument and more delicate differences than intentionally exaggerated here) that the velocity profile is indeed logarithmic but \emph{not} universal. In the relatively low-Reynolds example of Figure \ref{threevel}, \eqref{kinter} would give
\begin{itemize}
\item $\kappa=0.431$ for plane Couette flow,
\item $\kappa=0.386$ for plane duct flow,
\item $\kappa=0.350$ for circular pipe flow.
\end{itemize}

Route 3 is tempting, and we gave it some consideration before finding the more satisfactory solution that was described in \cite{PRL}. It corresponds to $\kappa=0.434$, about the same value as found for Couette flow above.

But in fact, none of these three alternatives turned out to be correct. They are here just to exemplify in how many different ways the same data may sometimes be interpreted.
In \cite{PRL} a two-term asymptotic expansion of the overlap layer was introduced as
\begin{multline}
u^+ = A_0\log(z^+)+A_1 g\Re_\tau^{-1}z^+ +B =\\
= A_0\log(Z)+A_1 g Z +A_0\log(\Re_\tau)+B,
\label{newovlap}
\end{multline}
where $A_0=\kappa^{-1}$, and $A_1$ is a new universal constant quantifying the effect of pressure gradient upon the turbulent velocity profile. The geometry parameter $g$, defined as $-hp_x/\tau_w=4h/D_H$, accounts for the different pressure gradient that corresponds to the same wall shear stress in each geometry, and takes on the values of $g=2$ for circular pipe flow, $g=1$ for pressure-driven flow in a plane duct and $g=0$ for turbulent Couette flow between countermoving plane walls. Equation \eqref{newovlap} collapses the three curves of Figure \ref{threevel} much better than \eqref{clasovlap}, as shown in Figure 2 of \cite{PRL}, and a universal description of the overlap layer is thus restored.

\subsection{Joint profile analysis}
\label{pairanalysis}
\begin{figure}
\includegraphics[width=\columnwidth]{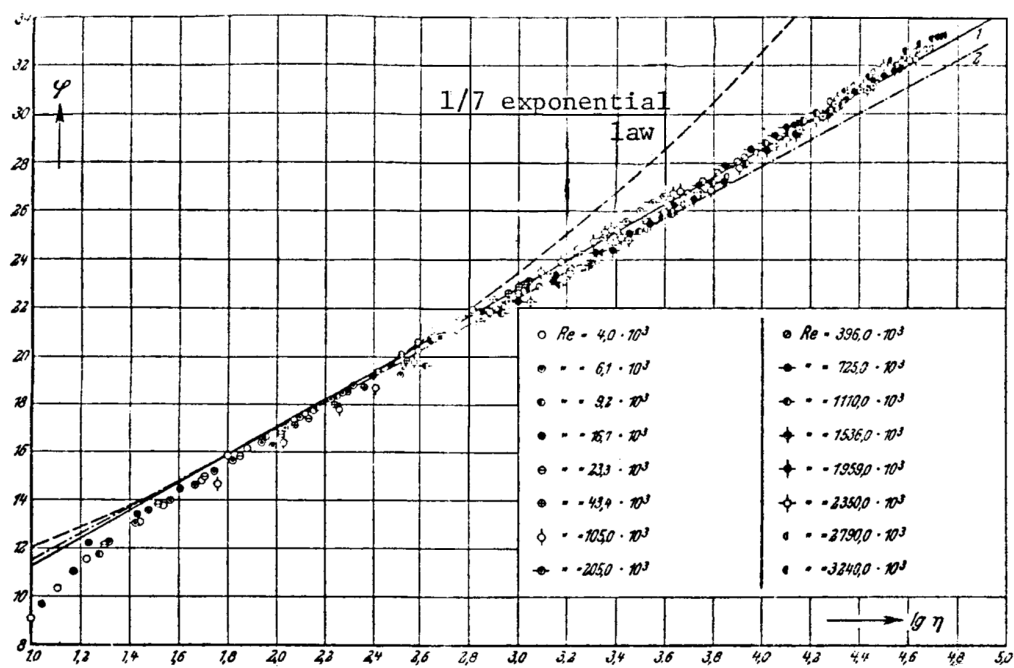}
\caption{a reproduction of Figure 24 of \cite{Nikuradse}. $\eta$, $\phi$ here are the same as $z^+$, $u^+$ in the text. Solid line: $\kappa=0.400$; dash-and-dot line: $\kappa=0.417$.}
\label{Nikupoints}
\end{figure}
\begin{figure}
\includegraphics[width=\columnwidth]{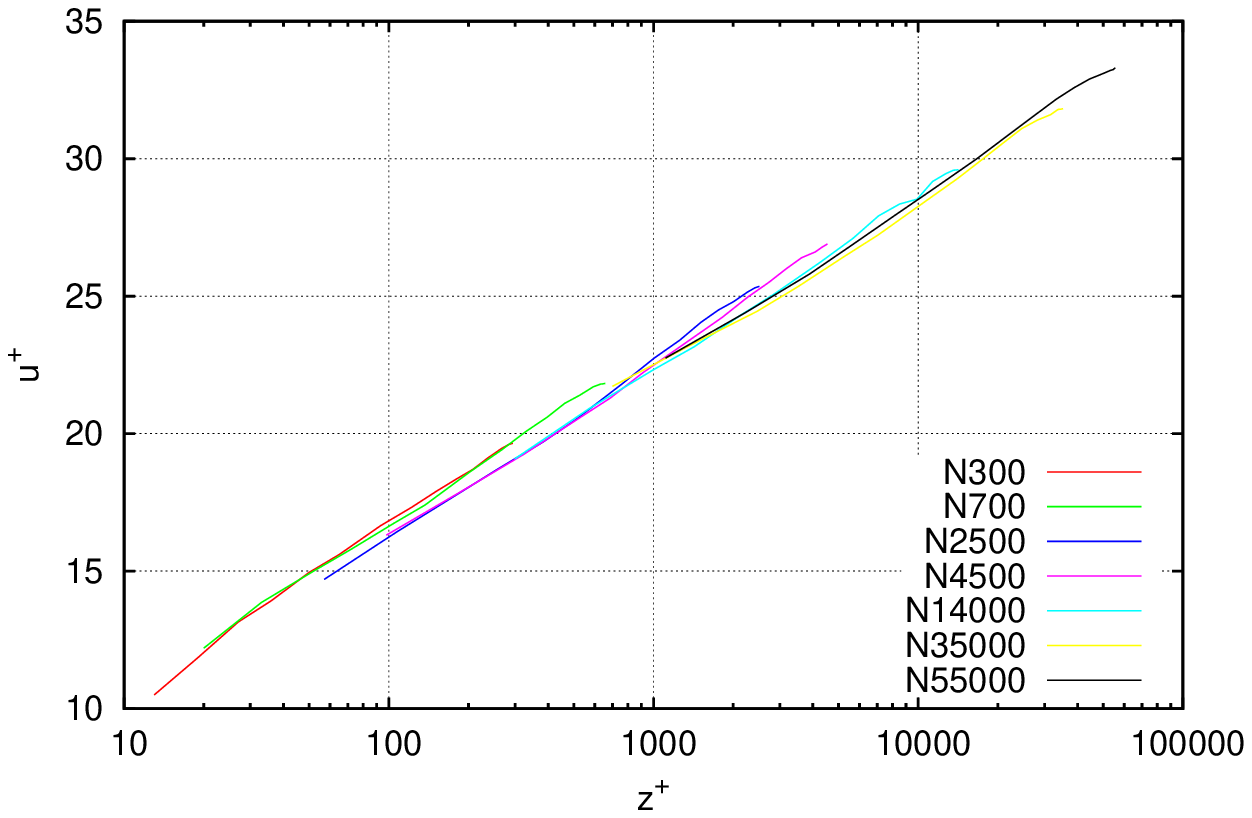}
\caption{Nikuradse's data replotted as lines. Each velocity profile is slightly slanted with respect to their common trend.}
\label{Nikuradselines}
\end{figure}

A digression must be made on the classical method used by experimentalists since Nikuradse \cite{Nikuradse} to estimate the theoretical parameters of the velocity profile \eqref{clasovlap}. This method is in fact none of the above, but rather to draw together in a single plot the velocity profiles obtained for different Reynolds numbers, and then graphically extract a common trend. The famous Figure 24 of Nikuradse \cite{Nikuradse}, reproduced as Figure \ref{Nikupoints} here, does so in a graphical format that uses marker symbols to denote experimental points; the unconscious effect is that the overall point cloud strikes the eye much more than the shape of individual velocity profiles. Nikuradse draws two straight lines, one corresponding to $\kappa=0.400$ that more or less cuts through the entire cloud and one corresponding to $\kappa=0.417$ that privileges measurements closer to the wall; the latter he believes to be more accurate because, as he explicitly remarks, his measurements near the centerline must be biased as the profiles do not end with a flat slope there. (We shall return to this observation later.) Nevertheless, the individual profiles are not visually distinguishable; but they become so if the same data are replotted as lines (Figure \ref{Nikuradselines}), which is easy to do because Nikuradse wisely provided his data in tabular form.
Had he recognized the importance of this information, he might have discovered the wake function twenty years before Coles.

In fact, from Figure \ref{Nikuradselines} we can observe that each individual curve is consistently slanted upwards with respect to their common trend, so that the slope estimated using a single Reynolds number can be nonnegligibly higher than the one estimated by joining all the data together. Their difference is due to the wake function $G(Z)$, as will be expanded upon later, and should be kept in mind when comparing data of different sources, possibly extracted by different methods. In fact, using more than one Reynolds number at the same time is a fortunate way to compensate for errors induced by  $G(Z)$.

On the other hand, figures such as Figure \ref{Nikupoints} can give the impression that all the point scatter is to be disregarded as measurement error. This way to interpret experimental data may have generated the legend that with the logarithmic law alone ``excellent agreement is obtained not only for points near the wall but for the whole range up to the axis of the pipe" \cite[p. 509]{Schlichting}, tantamount to saying, in more recent terminology, that the wake function $G(Z)$ is negligible for pipe flow. Instead, it can be seen from \eqref{newovlap} (and will be confirmed by examples in \S\ref{pipe}) that the wake function of pipe flow is twice as large as the one of plane parallel flow.

\subsection{Matched asymptotic expansions}
Equation \eqref{newovlap} is consistent with a pair of matched asymptotic expansions \textit{à la} van Dyke \cite{vanDyke} of the wall and defect layers, in which an asymptotic expansion in powers of $\Re_\tau^{-1}$ of the former is matched to a Taylor series in powers of $Z$ of the latter.

As documented in the reviews \cite{Buschmann,Panton05,Panton07}, several authors, starting from the observation that the ratio between the viscous length scale $l=\nu/u_\tau$ and the outer length scale $h$ in Prandtl-vonK\'arm\'an-Millikan theory is ${h^+}^{-1}=\Re_\tau^{-1}$, have proposed matched asymptotic expansions in powers of this parameter, modelled after the matched asymptotic expansions of laminar boundary-layer theory%
\footnote{
The choice of $\Re_\tau^{-1}$ as the expansion parameter is not obvious, though, because the ratio $u_\tau/U$ (which a posteriori turns out to be of the order of $(\log \Re_\tau)^{-1}$) is also a valid competitor for this role. $u_\tau/U$ was assumed as the basic expansion parameter of the boundary layer, for instance, by Mellor \cite{Mellor}. On the other hand, the choice of  $\Re_\tau^{-1}$ as the expansion parameter is made plausible by the general form of Millikan's matching condition and by the inclusion of Galilean invariance among its premises.}.
Typically an inner expansion for the wall layer $0 \le z^+\ll h^+$ is written as
\be
u^+= f_0(z^+)+{h^+}^{-1}f_1(z^+)+{h^+}^{-2}f_2(z^+)+\ldots,
\ee{wallexp}
and an outer expansion for the defect layer ${h^+}^{-1}\ll Z\le 1$ as
\be
u^+(h^+)-u^+= F_0(Z)+{h^+}^{-1}F_1(Z)+{h^+}^{-2}F_2(Z)+\ldots.
\ee{defectexp}
The derivative of each of the $f_i$ functions is then assumed to admit an asymptotic expansion in negative powers of $z^+$ for $z^+ \rightarrow \infty$ (producing a logarithmic term in $f_0$ and possibly higher terms) and each of the $F_i$ functions an expansion in positive powers of $Z$ for $Z\rightarrow 0$, leading to an order-by-order intermediate matching (\eg\ \cite{Afzal76,AY1973,BushF,Tennekes68,Yajnik70}) in the overlap layer.

It can be remarked that the assumption of an algebraic approach of the $f_i$'s towards their asymptotic behaviour (a series of negative powers of $z^+$ summed to as many positive powers as needed to fit the $F_i$'s) 
is at variance with the classical example of the laminar boundary layer, where this approach is exponentially fast. Whether an algebraic or an exponential behaviour better fits the turbulent flow is an interesting question we shall return to in \S\ref{wallfunction}. On the other hand, there is no prominent alternative to a regular expansion in positive powers of $Z$ as a suitable representation of the $F_i$'s.

Of course one has to be very careful when deducting an expansion like \eqref{wallexp} or \eqref{defectexp} from purely empirical data, even more careful than when deriving it from a differential equation, because the finite accuracy of the data will quickly be exceeded with increasing number of terms allowed.
For purposes of interpolation, an expression inspired by the uniform approximation \eqref{clasunif} may actually be more useful. This can be simply written as
\begin{multline}
u^+=f_0(z^+)+{h^+}^{-1}f_1(z^+)+{h^+}^{-2}f_2(z^+)+\ldots+\\
+G_0(Z)+{h^+}^{-1}G_1(Z)+{h^+}^{-2}G_2(Z)+\ldots
\label{unifexp}
\end{multline}

There is more than one, totally equivalent, possibility of defining the individual terms, because the intermediate overlap behaviour can indifferently be assigned to the $f$ or to the $G$ terms. For instance, the $A_1$ term in \eqref{newovlap} can be considered as pertaining to $G_0$, with the advantage that \eqref{clasunif} is unchanged at this order, or to $f_1$, doing which emphasizes that the pressure-gradient correction is of first order in the wall layer. 

In what follows we shall adopt the convention of associating the logarithmic and constant terms of \eqref{newovlap} to $f_0$, thus imposing that $G_0(0)=0$, and the linear term to $G_0$, thus imposing that $f_1$ = o$(z^+)$ for $z^+\rightarrow \infty$. Notice that equivalently we might have associated the linear term to $f_1$, but in that case \eqref{unifexp} to first order would have become
\begin{multline}
u^+=f_0(z^+)+{h^+}^{-1}\left[A_1 g z^+ + f_1(z^+)\right]+\ldots +H_0(Z)+\ldots
\label{varunif}
\end{multline}
with $H_0(z)$ defined as
\be
 H_0(Z)=G_0(Z)-A_1 g Z.
\ee{Hdef}
Both $H_0(Z)$ and $G_0(Z)$ vanish for $Z\rightarrow 0$ but the first does so much faster; either could be apt to be named ``law of the wake", but $G_0(Z)$ is more directly comparable to Coles' definition.

In the following sections we shall endeavour to extract the leading terms of \eqref{unifexp} from empirical (numerical and experimental) data, aiming at identifying the wake functions of the three parallel geometries of plane duct, circular pipe and Couette flow, and the universal wall function common to all three. %
Let us begin by examining each geometry in turn.

\section{Plane parallel duct flow}
\begin{figure}
\includegraphics[width=\columnwidth]{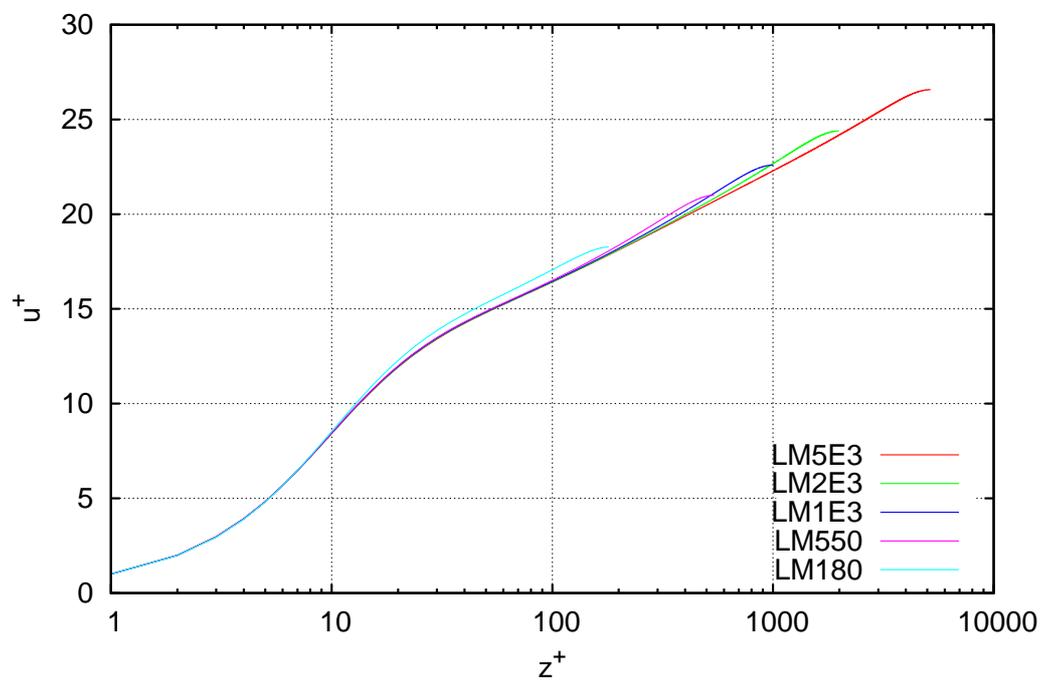}
\caption{Numerical velocity profiles from \cite{Moser}. Like Nikuradse's, these too are slanted and their individual slope is higher than their common trend.}
\label{planeorig}
\end{figure}

Flow driven by a pressure gradient between two infinite parallel plates (in short, plane-duct flow, or turbulent Poiseuille flow) is the simplest configuration where turbulent DNS (direct numerical simulation) can be performed, and the one for which the majority of DNS data are available, in contrast to flow in a parallel long pipe (turbulent Hagen-Poiseuille flow) for which the majority of experiments are performed. Since, as will be seen, the errors (intended as deviations of one data set from another) of DNS tend to be smaller we shall examine DNS data first, a sample of which is displayed in Figure \ref{planeorig}, with the purpose of separating functions $f(z^+)$ and $G(Z)$ from each other in equation \eqref{clasunif}. We shall start with the wake function which, being the smaller contribution, can afford a lower relative accuracy and still significantly improve the overall result.

\subsection{Estimating the wake function}
\label{wakeestim}
Examples of the wake function $G(Z)$ for pipes and channels are respectively given in Figs. 3,4 of Panton \cite{Panton07}, where they are obtained by difference between each measured, or numerically obtained, actual profile and a preassumed logarithmic law. This is the simplest way to obtain an estimate, and we shall return to it near the end of this paper, but the result is of course sensitive to the a-priori assumption of a log law with specific coefficients as Panton himself notes.

However, if just the uniform approximation \eqref{clasunif} is assumed to hold (for the time being, without any higher-order correction), and no log law is explicitly specified, there still is an unbiased way to identify its component functions $f(z^+)$ and $G(Z)$.
These functions can (in the absence of measurement or computation error) be extracted exactly from a pair of experimental or numerical profiles, measured on the same geometry at two different Reynolds numbers. Comparing the results obtained from more than one pair of Reynolds numbers then provides a test on the validity of \eqref{clasunif} itself.

Let $u^+_1$, $u^+_2$ denote two velocity profiles at Reynolds numbers $\Re_{\tau,1}=h^+_1$ and $\Re_{\tau,2}=h^+_2 > h^+_1$. According to \eqref{clasunif},
\begin{subequations}
\be
u^+_1(z^+)=f(z^+)+G(z^+/h^+_1)
\ee{u1u2a}
\be
u^+_2(z^+)=f(z^+)+G(z^+/h^+_2)
\ee{u1u2b}
\label{u1u2}
\end{subequations}

 This is a linear system of two equations in two unknowns from which we can easily eliminate $f(z^+)$ by subtraction. On defining
\be
\Delta G_{12}(z^+/h^+_1)=u^+_1(z^+)-u^+_2(z^+),
\ee{DeltaGdef}
we obtain
\be
\Delta G_{12}(Z)=G(Z)-G(mZ)
\ee{DeltaG}
with $m=h^+_1/h^+_2$. $\Delta G_{12}(Z)$ itself already constitutes a rough approximation of $G(Z)$ when $h_2 \gg h_1$, since by definition $G(0)=0$ and by continuity $G(mZ)\rightarrow 0$ when $m \rightarrow 0$.
But in fact, if \eqref{DeltaG} is regarded as a functional equation with $G(Z)$ as the unknown, we can just as well solve this functional equation exactly. Recursively applying the substitution $G(Z)\simeq\Delta G_{12}(Z)$ to the second term of \eqref{DeltaG} provides the tentative solution
\be
G(Z)=\sum_{n=0}^\infty \Delta G_{12}\left(m^n Z\right).
\ee{Gsol}
Since, by replacement of the independent variable,
\[
G(mZ)=\sum_{n=1}^\infty \Delta G_{12}\left(m^n Z\right),
\]
it is easily verified that \eqref{Gsol} is actually the exact solution of \eqref{DeltaG}. The summation converges quickly, and its numerical truncation to some finite $n$ is fairly straightforward. Once $G(Z)$ is calculated, $f(z^+)$ and its asymptotic, logarithmic or otherwise, behaviour can be immediately obtained (for $0\le z^+ \le h^+_2$) from \eqref{u1u2b} as
\be
f(z^+)=u^+_2(z^+)-G(z^+/h^+_2).
\ee{fsol}

\begin{figure}
\includegraphics[width=\columnwidth]{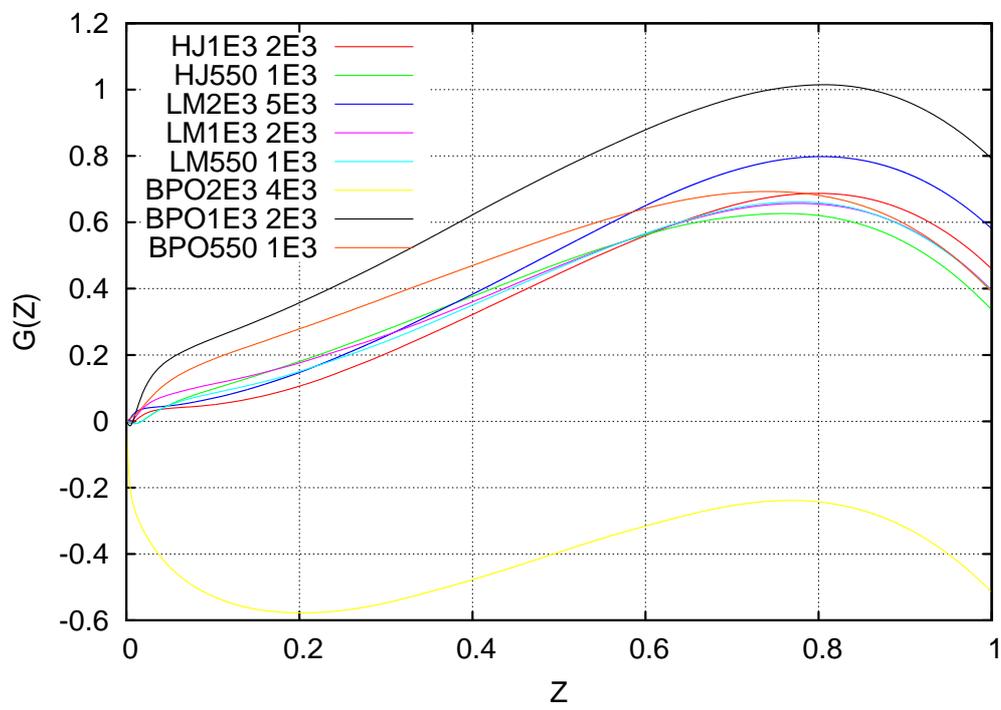}
\caption{Estimated law of the wake from several velocity profile pairs in plane parallel flow.}
\label{planewakeestim}
\end{figure}
A collection of curves obtained by applying (\ref{DeltaGdef},\ref{Gsol}) to the numerical data of \cite{Jimenez,Moser,Pirozzoli}  are exhibited in Figure \ref{planewakeestim}.
As may be seen, a general common shape of the defect-layer correction $G(Z)$ tends to be recognizable, as expected if \eqref{clasunif} is correct, but at the same time a large variability is present, not only between different Reynolds numbers but also between the results of different authors at the same Reynolds number, the data of \cite{Pirozzoli} being the most extreme\footnote{Notice that it is the subtraction of two profiles which amplifies the difference. On a standard $u^+$-$z^+$ plot the results of these same authors look hardly distinguishable. In this sense the present technique might also do double duty as a criterion of numerical convergence.}. This is a first indication that the precision (perhaps just the length of time averaging) of the numerical simulations available today still leaves room for improvement; it also hampers the distinction whether the difference between curves in Figure \ref{planewakeestim} is to be ascribed rather to higher-order corrections of the type \eqref{unifexp}, or to inaccuracy of the data. Nonetheless one can reckon that, if the data of \cite{Pirozzoli} are excluded, the maximum of $G(Z)$ in Figure \ref{planewakeestim} is of the order of $0.7$, and arrive at this number without any presumption of a logarithmic law or of the value of von Kármán's constant. Again with the exception of the data of \cite{Pirozzoli}, the vertical range of oscillation of $G(z)$ is of the order of $\pm 0.1$ on top of a velocity $u^+$ of the order of $20$, or about $\pm 0.5\%$. This can be easily ascribed to fluctuation of the finite-time numerical data averaging, as it would if \eqref{clasunif} were exact, although interestingly such fluctuation remains smooth across $Z$. %

Another feature of Figure \ref{planewakeestim} that may strike one's attention is that all curves end far from horizontal at $Z=1$, which clashes with Coles' idea of a half-wake (or mixing-layer) profile. This is the ``corner defect" whose discovery is ascribed by Panton \cite{Panton07} to Lewkowicz \cite{Lewkowicz}, although according to Musker \cite{Musker} it was noticed much earlier; with hindsight it is totally to be expected, because the $Z$-derivative of \eqref{clasovlap} at $Z=1$ is a constant $\kappa^{-1}$ independent of Reynolds number, and therefore $G'(1)=-\kappa^{-1}$ is required in order to have the velocity derivative $u_z$ vanish on the symmetry axis. The corner defect is of the same order of magnitude as the whole velocity defect, and remains so when $\Re_\tau\rightarrow\infty$, thus somewhat invalidating the traditional image of Coles' law of the wake as a monotonic quarter of a sine wave. (Although, it should be remarked, the latter was conceived with the boundary layer in mind, and Figure \ref{planewakeestim} might very roughly look like a sine extended over more than a quarter period.)

\begin{figure}
\includegraphics[width=\columnwidth]{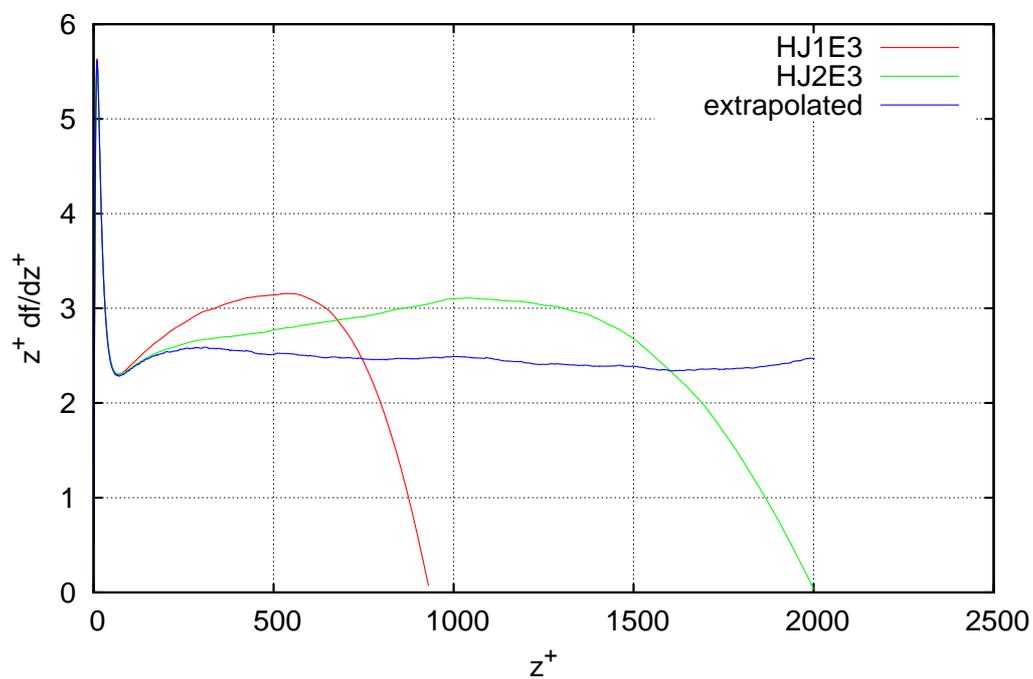}
\caption{Logarithmic derivative of the velocity profile for DNS at two Reynolds numbers, compared with the logarithmic derivative of the law of the wall as extrapolated from these same two profiles according to \eqref{fsol}.}
\label{ovlapderiv}
\end{figure}

Despite the relatively small correction provided by $G(Z)$ and the  large relative imprecision in its estimate, when $f(z^+)$ is extracted from one of these curves according to \eqref{fsol}, we obtain the third curve in Figure \ref{ovlapderiv}, with a very evident progress towards a range of nearly constant logarithmic derivative.
In particular, even if $z^+df/dz^+$ is not yet as constant as one would desire it to be in order to be able to estimate $\kappa$ precisely, it appears clearly for the first time that the constant range begins at $z^+\simeq 200$ (and is distinct from the minimum at $z^+\simeq 70$).
\begin{figure}
\includegraphics[width=\columnwidth]{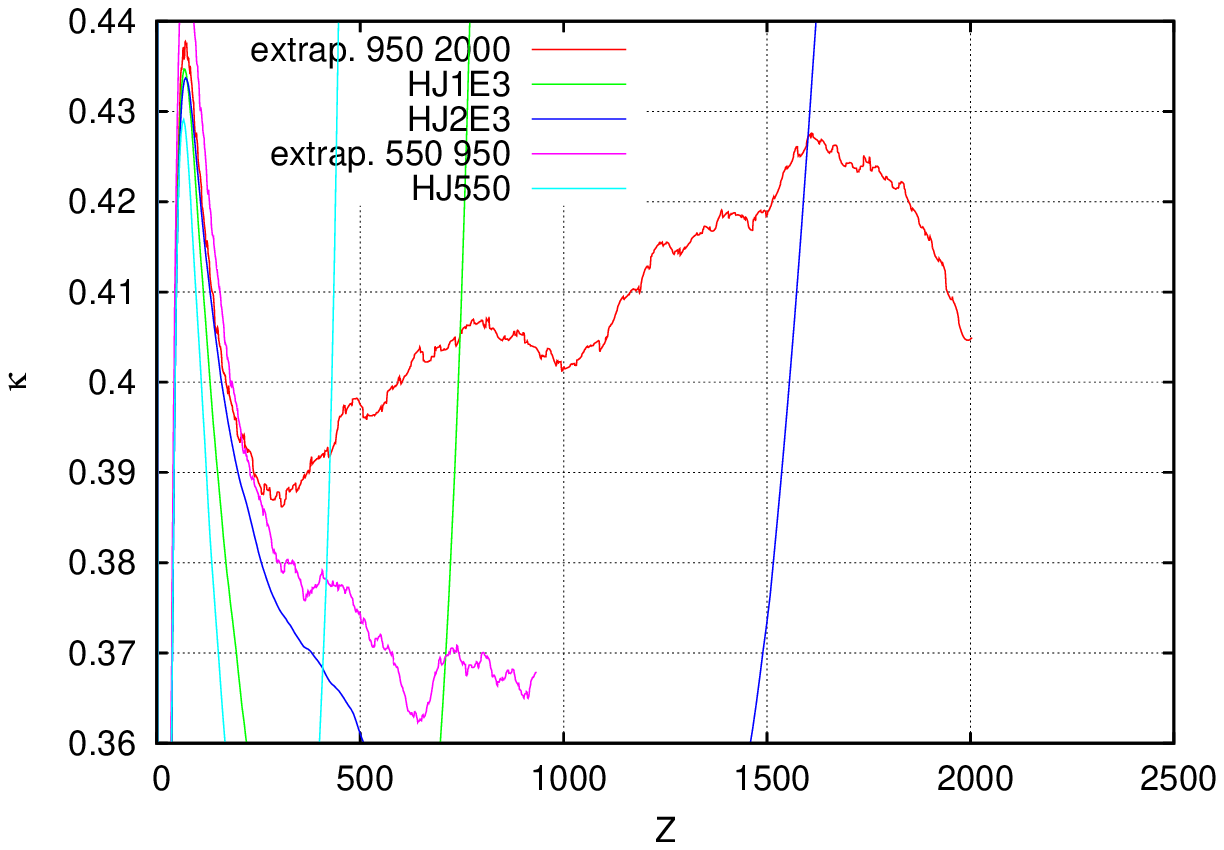}
\caption{A close-up of the extrapolated law of the wall (plotted as the reciprocal of its logarithmic derivative) highlights residual drift between different pairs.}
\label{planekappa}
\end{figure}
The result of Figure \ref{ovlapderiv} should eliminate any doubt that the Reynolds number of present numerical simulations is large enough to spot a logarithmic law; therefore, future research efforts should be directed at improving the accuracy of DNS more than at further increasing the Reynolds number.

Despite the strikingly different appearance of the extrapolated and original logarithmic-derivative profile, an enlargement of Figure \ref{ovlapderiv} shows that a systematic drift is still present. Figure \ref{planekappa} reports the reciprocal of the logarithmic derivative, which can be interpreted as a straight-out estimate of $\kappa$, in a zoomed-in subrange centered about $\kappa=0.4$. As can be seen, the precision of this plot is still insufficient to distinguish what the value of $\kappa$ is, except insofar as it is comprised between, say, $0.37$ and $0.42$. Nevertheless the error is not at random: we see a pattern of seemingly linear drift which in addition appears to change slope, and even sign, with Reynolds number (and, even if not shown in this figure, with data source). A possible interpretation and a way to overcome this drift will be given in next section.

\subsection{Eliminating the wake function}
\label{wakeelim}
Upon closer inspection of Figure \ref{planewakeestim}, one possible pitfall of the previous method comes to the eye: at small $Z$ the curves do not follow outer scaling. While this is not in itself unexpected, being the mark of higher-order corrections of the form \eqref{unifexp}, it may produce an error in the solution of \eqref{Gsol} which propagates from smaller to larger $Z$. In order to circumvent this error we can adopt a complementary approach which moves from large to small $Z$. This will bear some resemblance to the joint profile analysis of \S\ref{pairanalysis}, the one that underlies the data elaboration of many experiments; even if the composite expansion \eqref{clasunif} was formulated at a later time than the experimental method became established, it provides in fact a solid background for this method.

\begin{figure}
\includegraphics[width=\columnwidth]{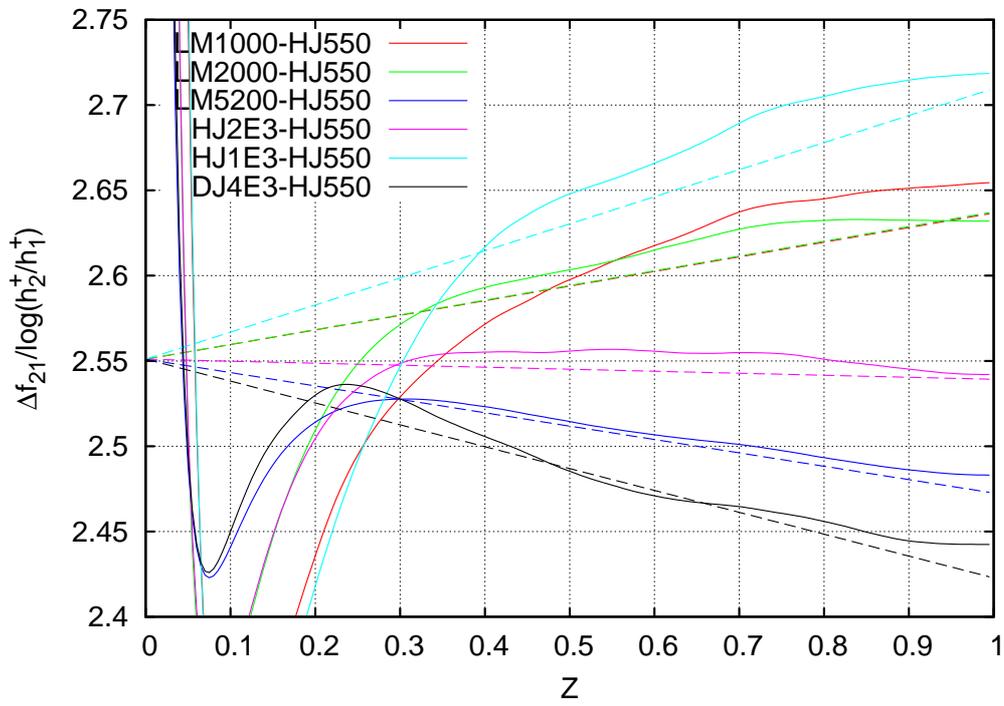}
\caption{Coefficient of the logarithmic law $A=\kappa^{-1}$ as estimated from eliminating the wake function according to \eqref{deltaf}.}
\label{rays0}
\end{figure}

If the difference of \eqref{u1u2a} and \eqref{u1u2b} is taken at equal $Z$, rather than at equal $z^+$, instead of eliminating $f(z^+)$ we obtain the complementary effect of eliminating $G$:
\be
\Delta f_{21}(Z)=u^+_2(h^+_2 Z)-u^+_1(h^+_1 Z)=f(h^+_2 Z)-f(h^+_1 Z).
\ee{deltaf}
In the defect layer, if $f \simeq A\log z^+ + B$ is assumed to hold with an unknown value of $A$, $\Delta f_{21}$ is expected to tend to a constant
\[
\Delta f_{21}(Z)\simeq A\log( h^+_2/h^+_1)\quad \text{for } 1/h^+_1\ll Z \le 1;
\]
therefore an alternate, easy way to estimate $A$, and von K\'arm\'an's constant $\kappa=A^{-1}$, is to plot the quantity $\Delta f_{21}/\log( h^+_2/h^+_1)$ as a function of $Z$ and look out for a constant range. A few such plots are displayed in Figure \ref{rays0} for various pairs of velocity profiles.

\begin{figure}
\includegraphics[width=\columnwidth]{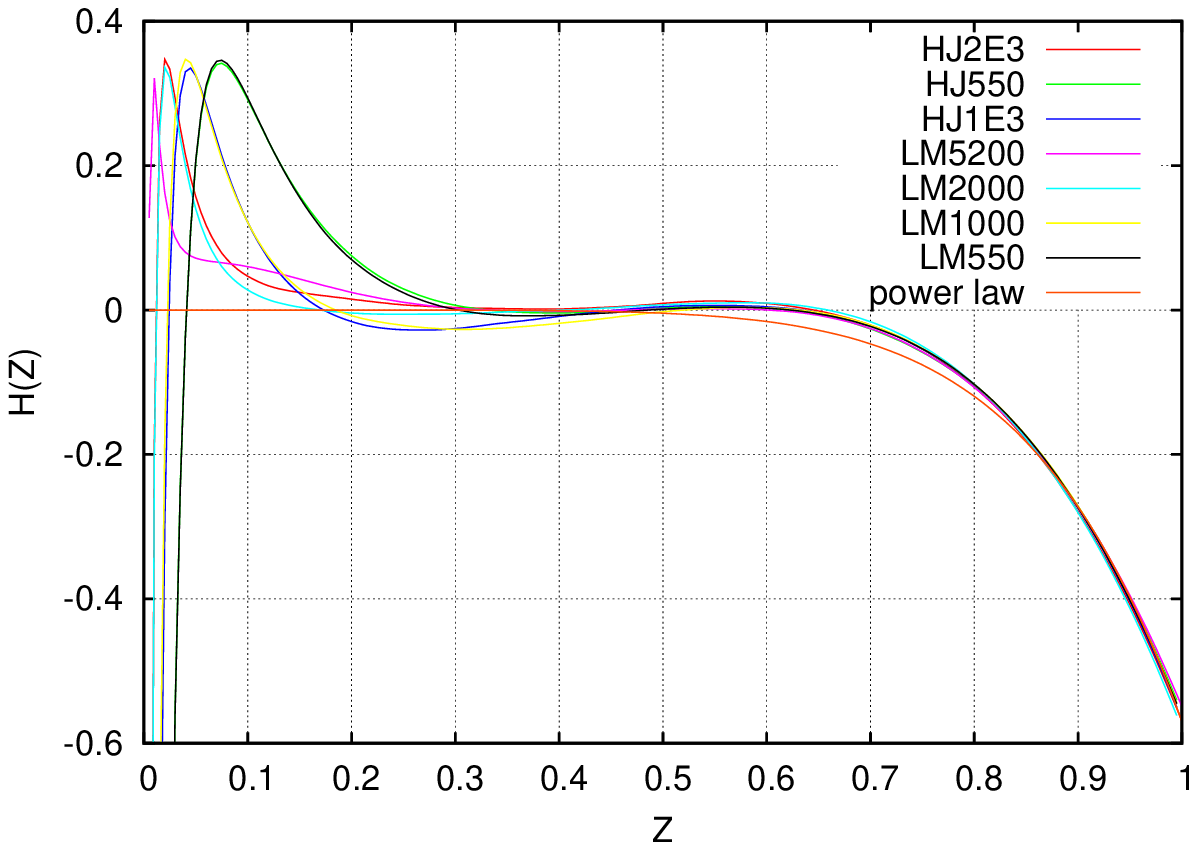}
\caption{Wake corner defect and its power-law interpolation.}
\label{Hwake}
\end{figure}

Just as was the case in Figure \ref{planekappa}, the estimate is consistent with the presence of a range where the plotted quantity is fairly constant (notice that the vertical scale is quite enlarged), but does not exhibit the level of precision required to locate the value of von K\'arm\'an's constant precisely. As shown by the overlaid dashed lines, however, this figure displays a remarkable pattern: once inside the defect layer, all curves tend to align along straight lines that radiate from a common point located at $A_0 = 2.55$, or $\kappa=0.392$.

Despite the straight-line pattern in Figure \ref{rays0} is very evident, its origin is not so, but it can be remarked that, whereas a single horizontal line would be expected on the basis of \eqref{clasunif}, \eqref{newovlap}, and the universality of the $A_0$ and $A_1$ constants, the observed behaviour becomes predictable if a common value of $A_0$ but a different value of $A_1$ is used to fit each curve. The best-fit values of $A_1$ are listed in Table \ref{A1}.
\begin{table}
\caption{Coefficient of the linear velocity correction as fitted to individual velocity profiles.}
\label{A1}
\centering
\begin{tabular}{llll}
Reference & $\Re_\tau$ & $A_0$ & $A_1$ \\
\hline
\cite{Jimenez} & 547 & 1/0.392 & 1.01 \\
\cite{Jimenez} & 934 & 1/0.392 & 1.10 \\
\cite{Jimenez} & 2003 & 1/0.392 & 1.00 \\
\cite{Jimenez4000} & 4179 & 1/0.392 & 0.76 \\
\cite{Moser} & 543 & 1/0.392 & 1.05 \\
\cite{Moser} & 1000 & 1/0.392 & 1.06 \\
\cite{Moser} & 1995 & 1/0.392 & 1.13 \\
\cite{Moser} & 5186 & 1/0.392 & 0.84 \\
\end{tabular}
\end{table}

A hint at the origin of this variability in $A_1$ comes from the observation that there are differences between results of different authors at the same Reynolds number, and the dependence on Reynolds number itself looks somewhat erratic. These are indications that variations in $A_1$ are more likely ascribed to numerical error than to the physics of turbulence. Why the error should settle along straight lines is not obvious, but we can offer the following speculation: the slowest decaying mode that perturbs the time averaging is tantamount to a modification in $A_1$, or in the interaction between imposed pressure gradient and imposed flow rate. That this or another be the correct interpretation, somewhat astonishingly numerical error produces straight lines in this plot. In what follows we shall proceed based on the conjecture that
\begin{quotation}
\it
\noindent Equation \eqref{newovlap} is the correct expression for the overlap layer, our present best approximation of $\kappa$ is $0.392$, and different numerical simulations err in their estimate of $A_1$ according to Table \ref{A1}.
\end{quotation}
This conjecture is tantamount to an, admittedly bold, estimate of the numerical error in order to subtract it, and the reader may or may not want to accept its implications, but some intriguing deductions can be drawn as will soon appear. Well aware that only a set of more accurate simulations will be able to clarify the matter in the future, we shall present the results of such deductions both with a single common value of $A_1$ and with the values of $A_1$ taken from Table \ref{A1}.

Whatever the origin of the pattern in Figure \ref{rays0}, the value of $A_0=1/0.392$ is bound to play a special role, and as will be seen all our further tests in this paper are consistent with it being our present best approximation of $\kappa$. It can also be noted that the estimates of $A_1$ in Table \ref{A1} tend to cluster about unity, and most importantly, the pair of profiles that in Figure \ref{rays0} turn up to have a difference with almost flat slope, \ie\ the best adherence to a logarithmic law (HJ550 and HJ2E3), have in Table \ref{A1} a value of $A_1$ very close to unity. All this leads us to further infer that the correct value may be exactly $A_1=1$, as will also be confirmed by the comparison with cylindrical pipe flow in \S\ref{pipe}. The values of $\kappa=0.392$, $A_1=1$ and $B=4.48$, giving rise to the logarithmic law
\be
u^+=\log(z^+)/0.392+4.48,
\ee{loglaw}
were reported in \cite{PRL} as our present best estimates.

An immediate consequence of allowing a different value of $A_1$ for each DNS %
is that a very precise identification of the corner-defect function $H(Z)$ of \eqref{Hdef} ensues. This is brought out in Figure \ref{Hwake}, where \eqref{newovlap} is subtracted from each profile with $A_0=1/0.392$ and $A_1$ taken from Table \ref{A1}. The resulting estimates of $H(Z)$ are nearly identical (compare the accord in Figure \ref{Hwake} to the imprecision in Figure \ref{planewakeestim}), and for the purpose of interpolation can be fit by the power-law
\be
H(Z)\simeq-0.57 Z^7 .
\ee{Hintrp}

It follows from \eqref{Hdef} that, within this approximation, Coles' wake function $G(Z)$ can be usefully expressed as
\be
G(Z)\simeq A_1 Z-0.57 Z^7
\ee{Gintrp}
Once again, this function has a different shape than traditionally has been associated with Coles' law of the wake (a quarter sine wave with flat slope both at $Z=0$ and at $Z=1$, or a similarly shaped cubic polynomial). It must be remarked, however, that Coles \cite{Coles} drew this shape with the boundary layer in mind, which might behave differently from plane-duct flow in this respect.

\begin{figure}
\includegraphics[width=\columnwidth]{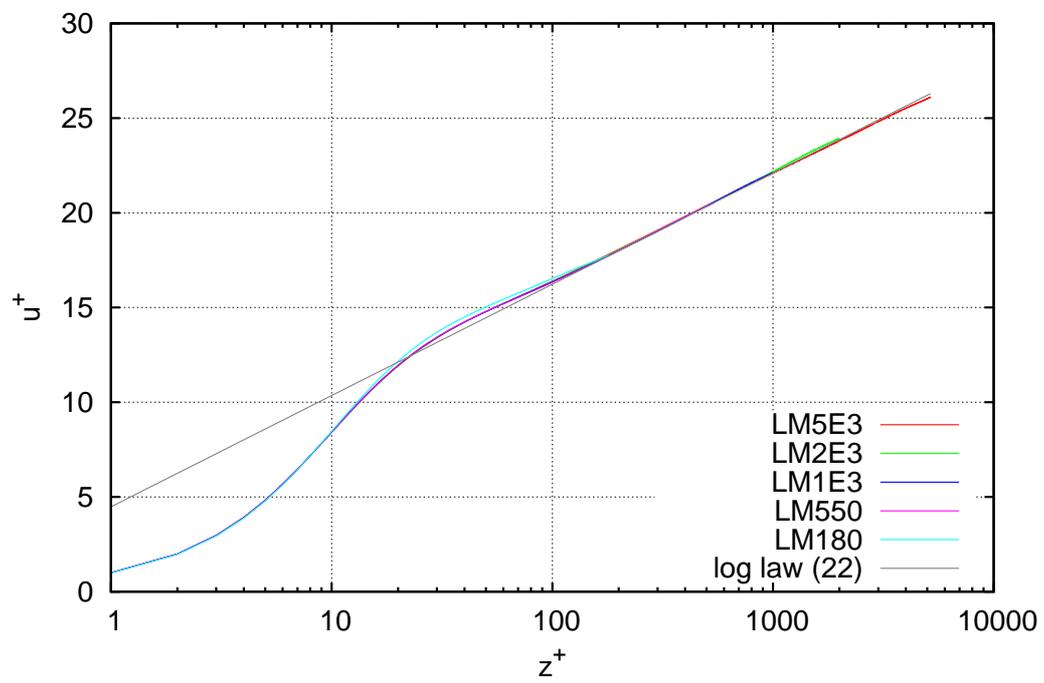}
\caption{Same velocity profiles as in Figure \ref{planeorig} after subtracting the wake function \eqref{Gintrp} with coefficient $A_1=1$.}
\label{plane}
\end{figure}
\begin{figure}
\includegraphics[width=\columnwidth]{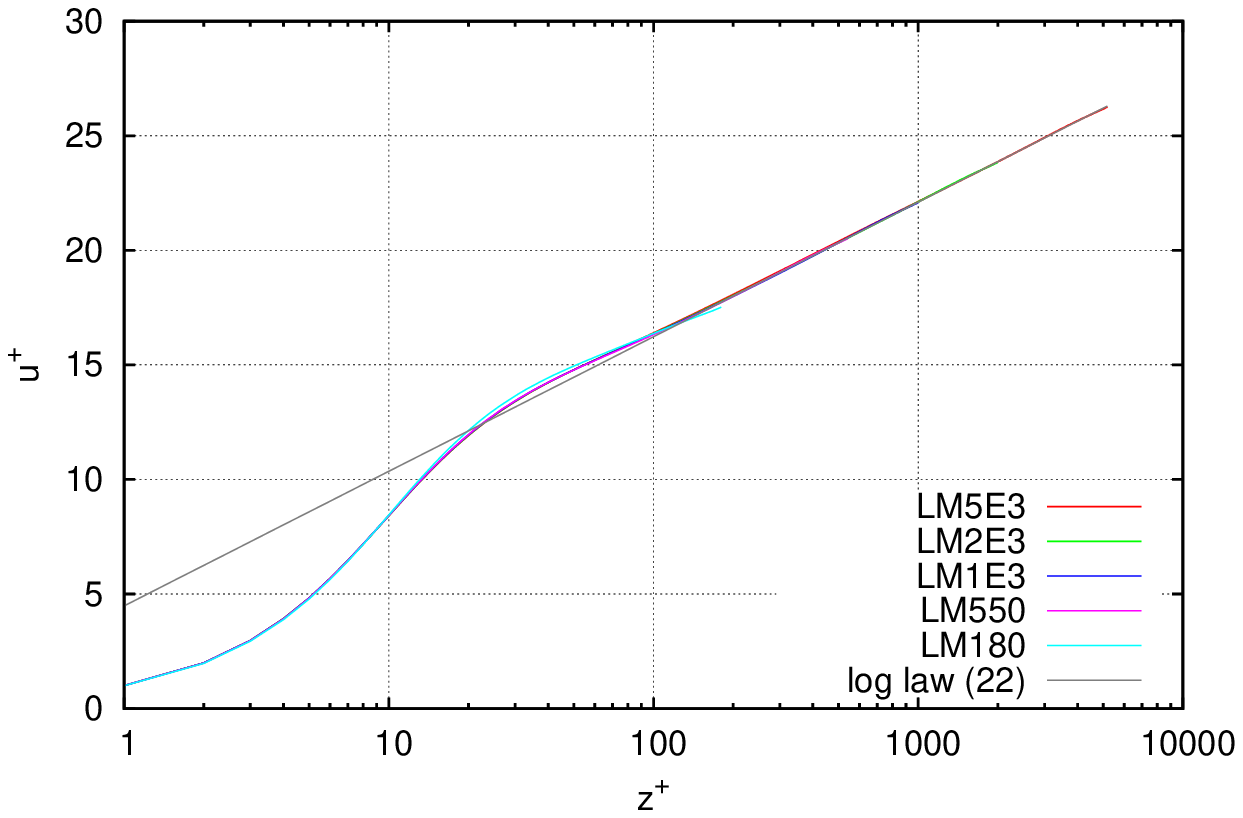}
\caption{Same velocity profiles as in Figure \ref{planeorig} after subtracting the wake function \eqref{Gintrp} with coefficient $A_1$ from Table \ref{A1}.}
\label{planeTable}
\end{figure}
Eventually, when the complete wake function \eqref{Gintrp} is subtracted from the velocity profile so as to extract the wall function $f(z^+)$, the outcome is Figure \ref{plane} with $A_1=1$ or Figure \ref{planeTable} with $A_1$ taken from Table \ref{A1}, in either case a very substantial improvement over Figure \ref{planeorig} as far as agreement to a logarithmic law is concerned.

\section{Pipe flow}
\label{pipe}
Numerical velocity profiles for circular pipe flow are less abundant and extend to a smaller Reynolds number than for flow in a plane duct; in exchange the majority of experimental data pertain to this geometry.

\subsection{DNS data}
\begin{figure}
\includegraphics[width=\columnwidth]{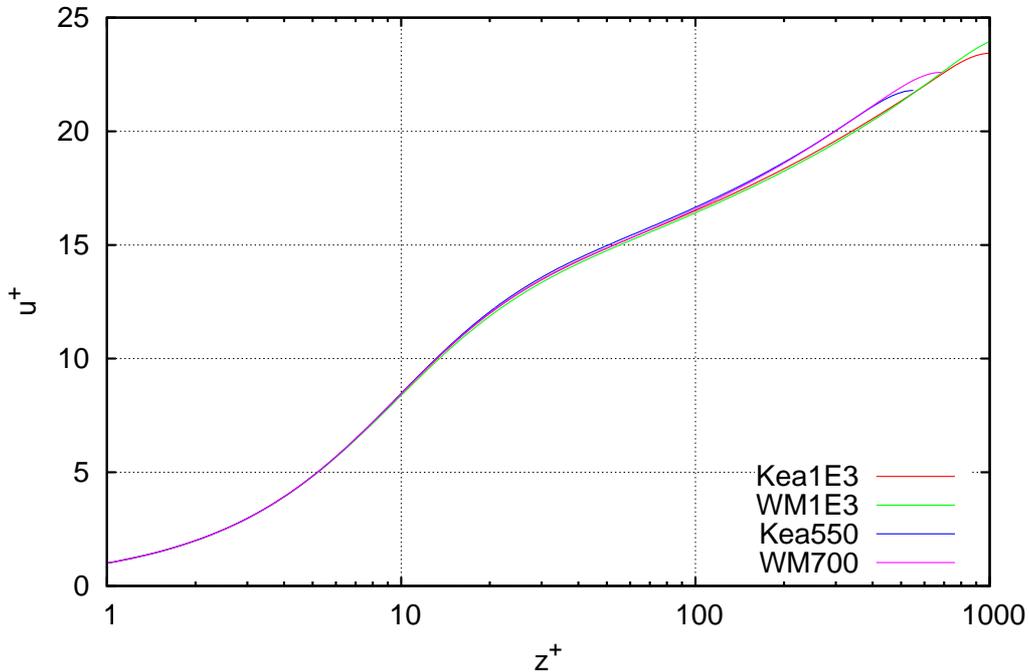}
\caption{Numerical velocity profiles for pipe flow from \cite{Schlatter,Moin}. These are even more distant from a logarithmic behaviour than those in Figure \ref{planeorig}.}
\label{pipeorig}
\end{figure}
Figure \ref{pipeorig} shows the unmanipulated velocity profiles of El Khouri \etal\ \cite{Schlatter} and Wu and Moin \cite{Moin} up to $Re_\tau\gtrsim 1000$. As can be seen, they depart even more pronouncedly from a straight line than the plane-duct profiles did in Figure \ref{logderiv}.

\begin{figure}
\includegraphics[width=\columnwidth]{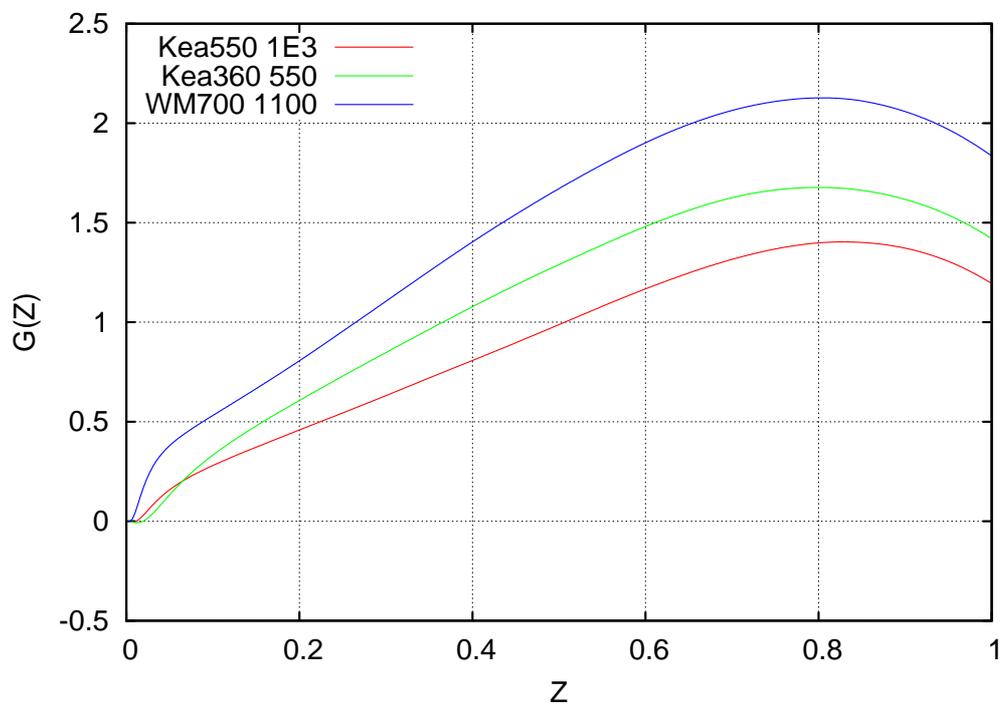}
\caption{Wake estimation method of \S\ref{wakeestim} as applied to pipe flow DNS.}
\label{pipewakeestim}
\end{figure}

When the wake-extraction technique of \S\ref{wakeestim} is applied to these profiles, the result is Figure \ref{pipewakeestim}.
Here the variability is larger than in Figure \ref{planewakeestim}, mostly due to the lower range of Reynolds number but also probably to a shorter period of time averaging; nonetheless a very similar shape of the wake function can be spotted, except that the vertical scale is doubled. As was explained in \cite{PRL}, this is no accident and can be correlated to the fact that the pressure gradient is also doubled in this geometry. Here just as in Figure \ref{planewakeestim}, an initial kink arises on the wall scale; this kink requires a word of caution since it might be responsible, through error propagation in \eqref{Gsol}, of a possibly artificial vertical shift of the whole curve. 

\begin{figure}
\includegraphics[width=\columnwidth]{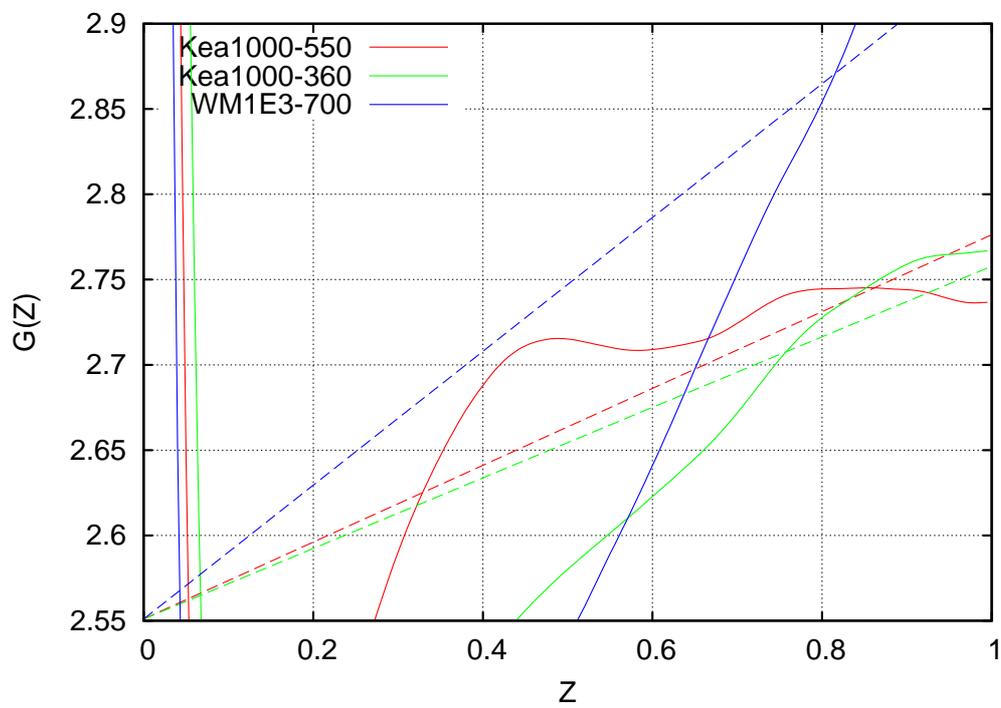}
\caption{Wake elimination method of \S\ref{wakeelim} as applied to pipe flow DNS.}
\label{piperays0}
\end{figure}
\begin{figure}
\includegraphics[width=\columnwidth]{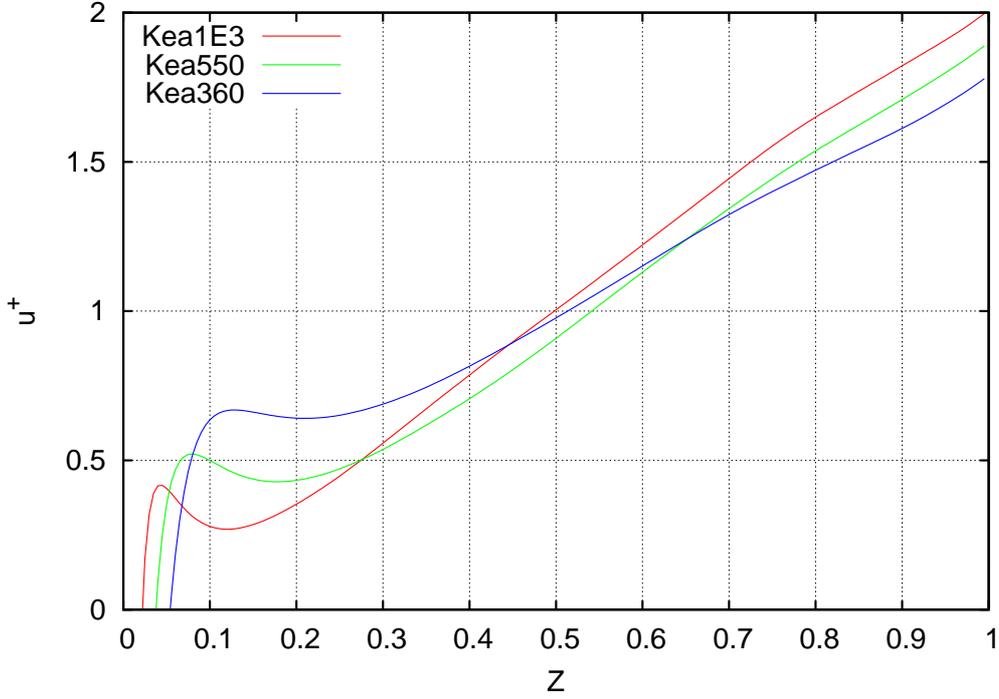}
\caption{Pipe velocity profiles, after subtracting the logarithmic law and corner-defect function, approach a straight line with slope $gA_1=2$.}
\label{verifpipe}
\end{figure}

When the wake-elimination technique of \S\ref{wakeelim} is applied to pipe flow DNS results, the result is Figure \ref{piperays0}. The pattern of straight lines looks forced upon this figure, again an indication of the lower range of Reynolds numbers but also probably of a generally larger residual numerical error, and  it cannot honestly be spotted without a comparison with Figure \ref{rays0}. Nonetheless, nontrivially this plot is \emph{compatible} with a straight-line pattern emanating from the \emph{same value of $\kappa$} as the plane profiles. This becomes fairly evident if, instead of taking differences between profile pairs, we accept as an ansatz to be proved %
 that the log law coincides with the one found for plane flow, including its coefficients $\kappa=0.392$ and $B=4.48$, and, in Figure \ref{verifpipe}, subtract from each profile the log law \eqref{loglaw} and the corner-defect function.
There should now be hardly any doubt that the difference approaches a straight line with a slope close to $gA_1=2$ ($g$ being the geometry factor defined in \S\ref{pgcorr}, $g=2$ for pipe flow). The best-fit slopes are reported in Table \ref{pipeA1}.
\begin{table}
\caption{Best fit coefficient of the linear velocity correction for pipe flow.}
\label{pipeA1}
\centering
\begin{tabular}{llll}
Reference & $\Re_\tau$ & $A_0$ & $gA_1$ \\
\hline
\cite{Schlatter} & 361 & 1/0.392 & 1.83 \\
\cite{Schlatter} & 550 & 1/0.392 & 1.91 \\
\cite{Schlatter} & 999 & 1/0.392 & 2.04 \\
\cite{Moin} & 685 & 1/0.392 & 2.22 \\
\cite{Moin} & 1142 & 1/0.392 & 2.43 \\
\end{tabular}
\end{table}
As to the corner-defect function,
\begin{figure}
\includegraphics[width=\columnwidth]{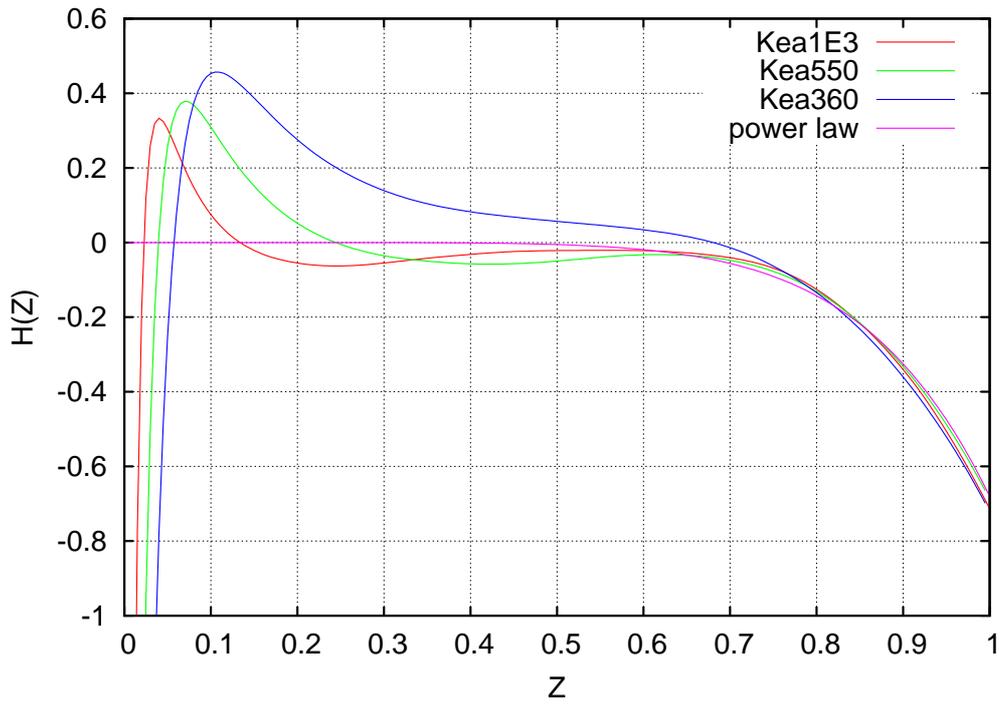}
\caption{Corner-defect function for pipe flow.}
\label{wakefuncp}
\end{figure}
this is shown in Figure \ref{wakefuncp} and can be approximated as
\be
H(Z)= - 0.67 Z^7 .
\ee{pwakeH}
When eventually the complete wake function
\be
G(Z)=gA_1 Z - 0.67 Z^7
\ee{pipeG}
is subtracted from each velocity profile, the estimate of the universal law of the wall in Figure \ref{pipef} ensues, which should be compared to the untreated Figure \ref{pipeorig} as far as adherence to a logarithmic law is concerned.
\begin{figure}
\includegraphics[width=\columnwidth]{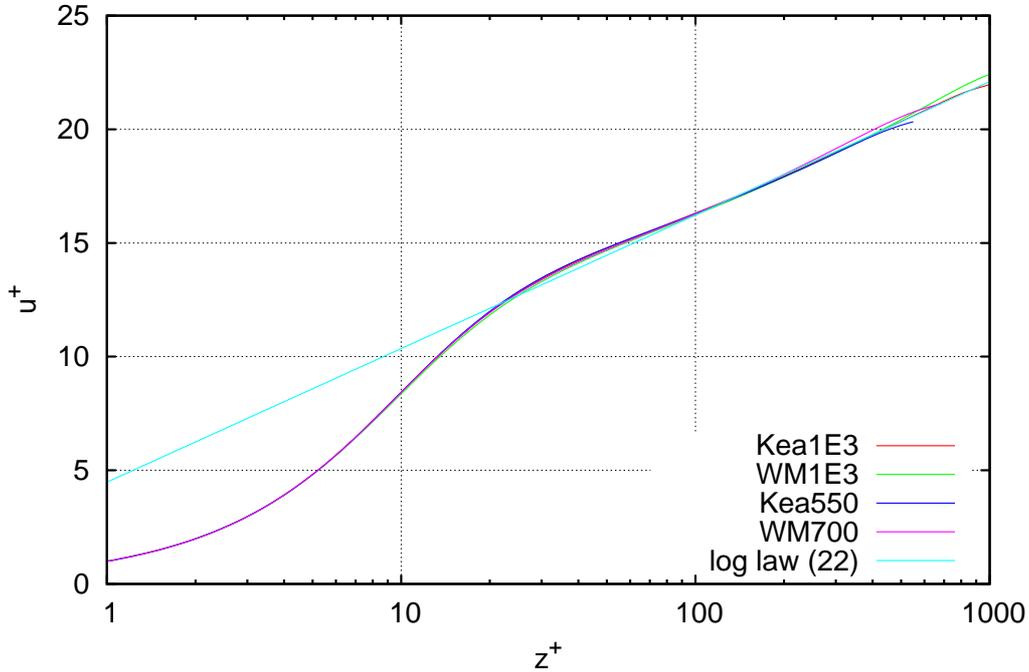}
\caption{Same velocity profiles as in Figure \ref{pipeorig} after subtracting the wake function \eqref{pipeG} with coefficient $gA_1=2$.}
\label{pipef}
\end{figure}

\subsection{Experimental data}
Experiments on turbulent pipe flow extend up to a Reynolds number $\Re_\tau\simeq 10^5$, which is a hundred times larger than the largest value yet achieved in DNS. On the other hand they present generally larger variability, and the higher the Reynolds number the worse they resolve the wall region for obvious instrument-size constraints. Lack of the wall region renders the method of \S\ref{wakeestim} inapplicable, since \eqref{Gsol} needs a reasonably good interpolation there.
\begin{figure}
\includegraphics[width=\columnwidth]{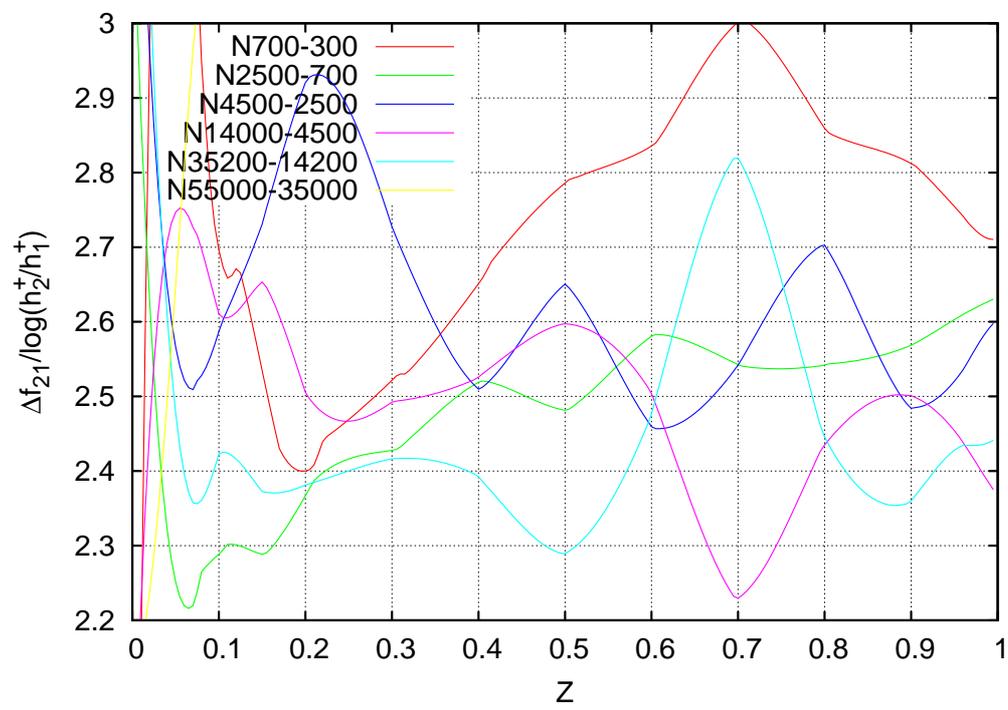}
\caption{Wake elimination method of \S\ref{wakeelim} as applied to Nikuradse's pipe flow experiments.}
\label{Nikurays}
\end{figure}
The wake-elimination method of \S\ref{wakeelim} does not suffer from this limitation, since profile differences are only taken in the defect layer, but error amplification by such differences renders the result, visible in Figure \ref{Nikurays}, of little quantitative value (even though the barycenter of the whole figure is not far from the by now expected $A_0=2.55$).

\begin{figure}
\includegraphics[width=\columnwidth]{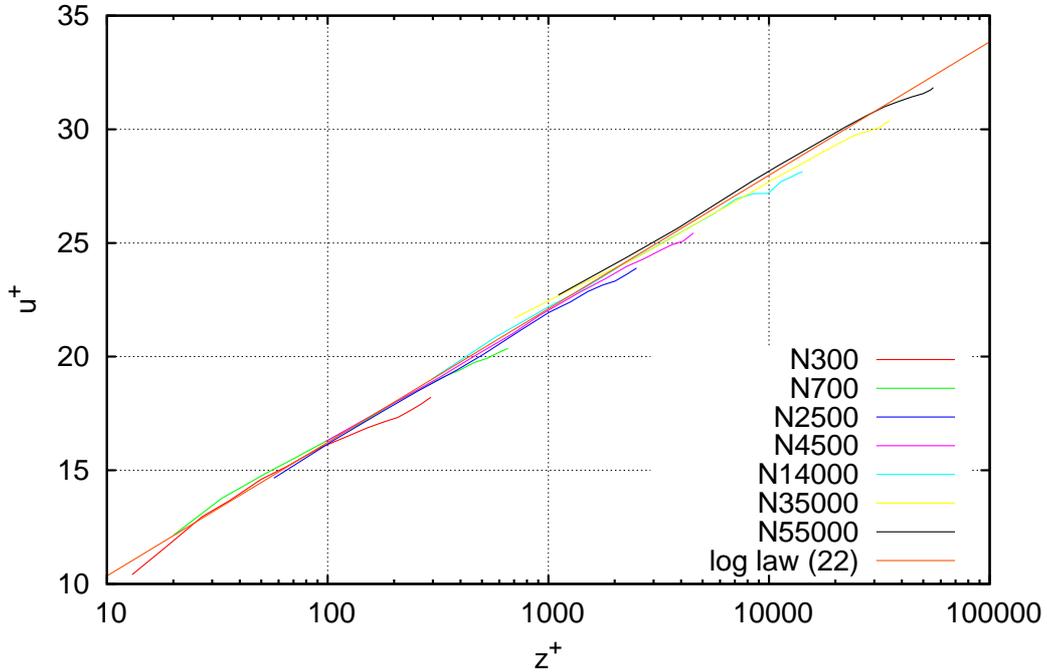}
\caption{Same velocity profiles as in Figure \ref{Nikuradselines} after subtracting their linear pressure-gradient correction with coefficient $gA_1=2$.}
\label{Nikuf}
\end{figure}

On the other hand, if again the same wake function \eqref{pipeG} is subtracted as obtained from previous cases, Nikuradse's original data in Figure \ref{Nikuradselines} become those in Figure \ref{Nikuf},
and exhibit a close fit to the the logarithmic law \eqref{loglaw} with coefficients $\kappa=0.392$ and $B=4.48$; a fit which is even more convincing insofar as these coefficients were extrapolated from a Reynolds number a hundred times smaller. In fact, at closer inspection the individual velocity profiles in Figure \ref{Nikuf} are now no longer slanted and much more parallel to the log law, the main discrepancy being a deviation downwards in the rightmost part of each curve corresponding to the corner defect. Nikuradse himself comments that his measurements are not completely reliable near the center.%

\begin{figure}
\includegraphics[width=\columnwidth]{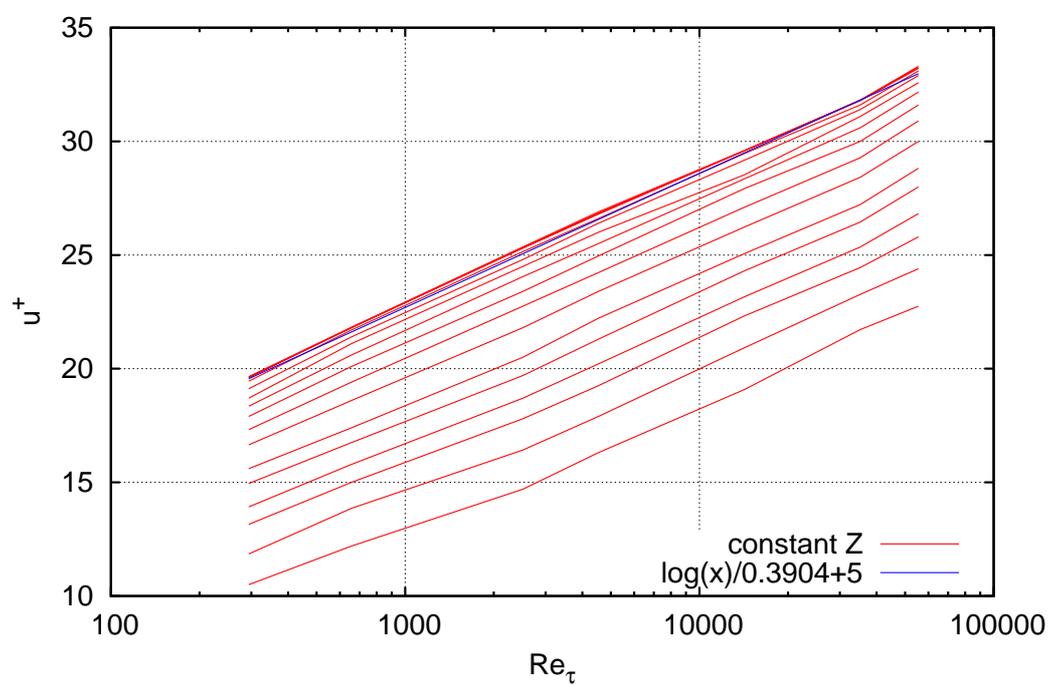}
\caption{$u^+$ as a function of $\Re_\tau$ for constant $Z$.}
\label{NikZ}
\end{figure}

Actually Nikuradse \cite{Nikuradse} concluded that his most reliable estimate was $\kappa=0.417$, but he narrowly missed a more accurate result with his Figure 18, where he looks for the dependence of velocity on Reynolds number at a constant position rather than the opposite. Had he plotted $u/u_\tau$ as a function of $\Re_\tau$ for constant $Z$, instead of $u/U$ as a function of $\Re_\text{bulk}$ for constant $Z$, he would have implicitly eliminated the contribution of the wake function $G(Z)$ and obtained the present Figure \ref{NikZ}, where the best fit slope corresponds to $\kappa=0.3904$.

\begin{figure}
\includegraphics[width=\columnwidth]{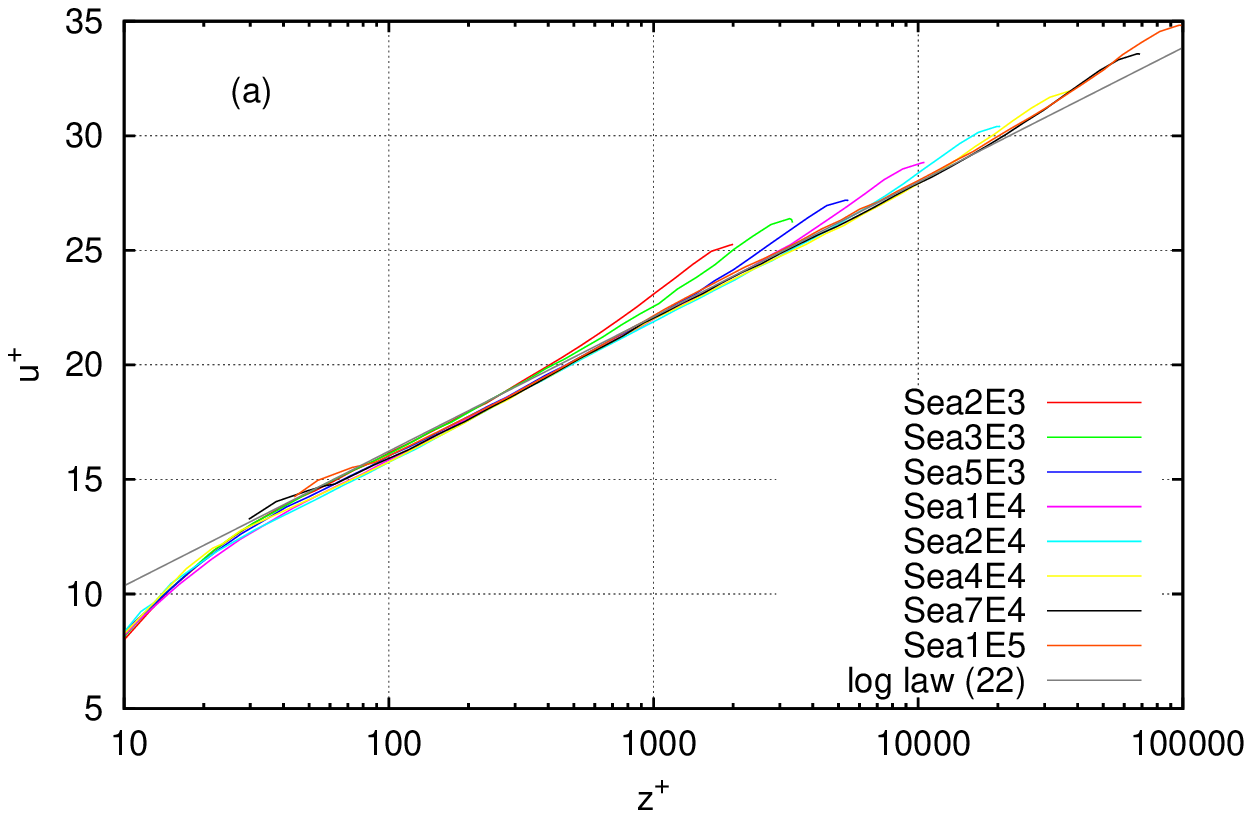}

\includegraphics[width=\columnwidth]{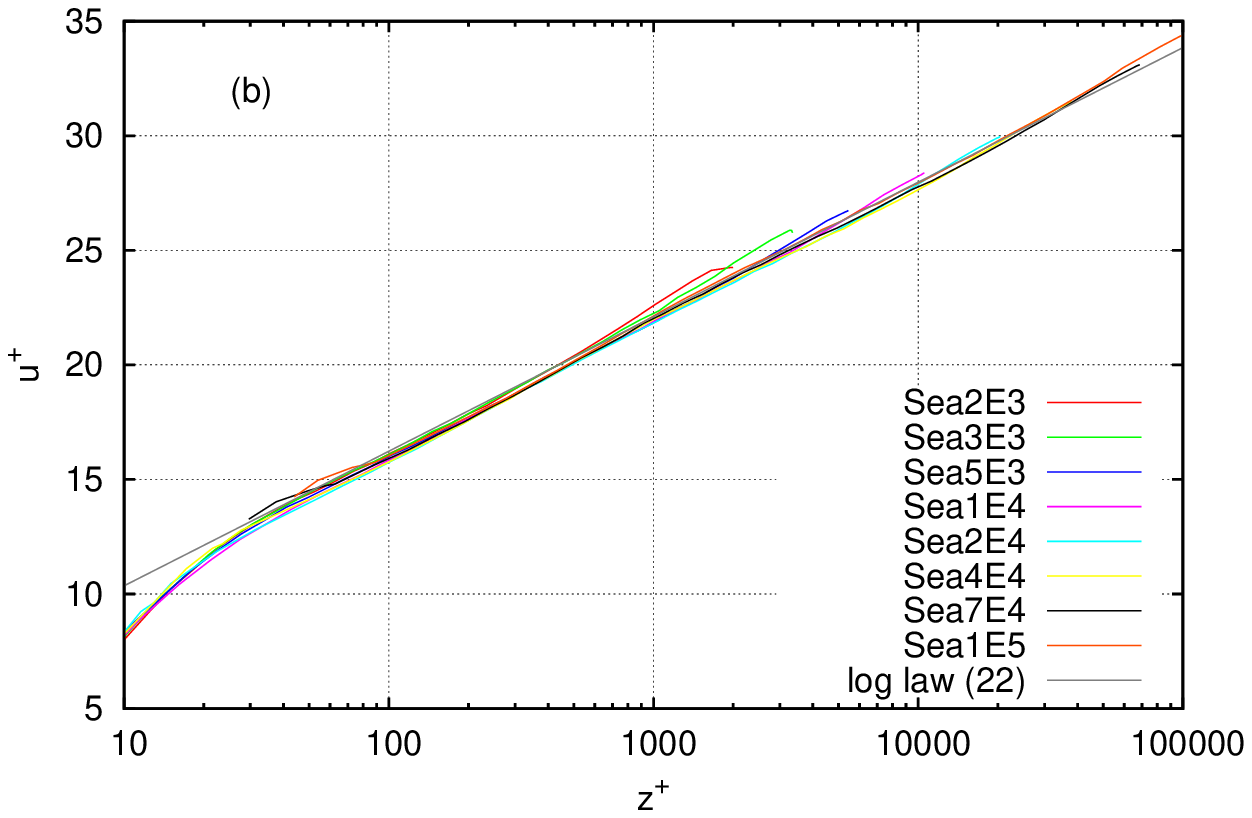}
\caption{Superpipe velocity profiles from \cite{Smits}, (a) before and (b) after subtraction of the wake contribution.}
\label{superpipe}
\end{figure}

Similar considerations apply to the Superpipe experiment of \cite{Smits}, whose raw velocity profiles (actually, ``raw" after the fine corrections described in \cite{Smits}) are reported in Figure \ref{superpipe}a. The upward slant is even more evident than in Nikuradse's.

When the wake function \eqref{pipeG} is subtracted, these become the profiles displayed in Figure \ref{superpipe}b, with again a sizeable progress towards the same logarithmic law \eqref{loglaw}. The main discrepancy is this time a slight residual upwards deviation in the rightmost part of each curve, which when compared to Nikuradse's excessive downwards deviation corroborates the idea that such residual deviations, in either case smaller than deviations of the original data, might actually be measurement errors.

Taken all together, the different experimental curves tend to satisfactorily tighten up onto the logarithmic law \eqref{loglaw} when the wake function, including the pressure-gradient correction, is subtraced. %
It must barely be reminded that this is not a logarithmic law fitted to these particular data, but one having the same numerical coefficients (both $A_0$ and $B$) as in all our other plots.

\section{Couette flow}
\label{couette}
Turbulent Couette flow has been much less intensely investigated than pipe or plane-duct flow, and we have the DNS of \cite{Pirozzoli} as our only source. Nevertheless it is an essential contribution to our comparison, as it completes the picture with a third value of the geometry factor $g$. ($g=2$ for pipe flow, $g=1$ for plane-duct flow, $g=0$ for turbulent Couette flow.)

\begin{figure}
\includegraphics[width=\columnwidth]{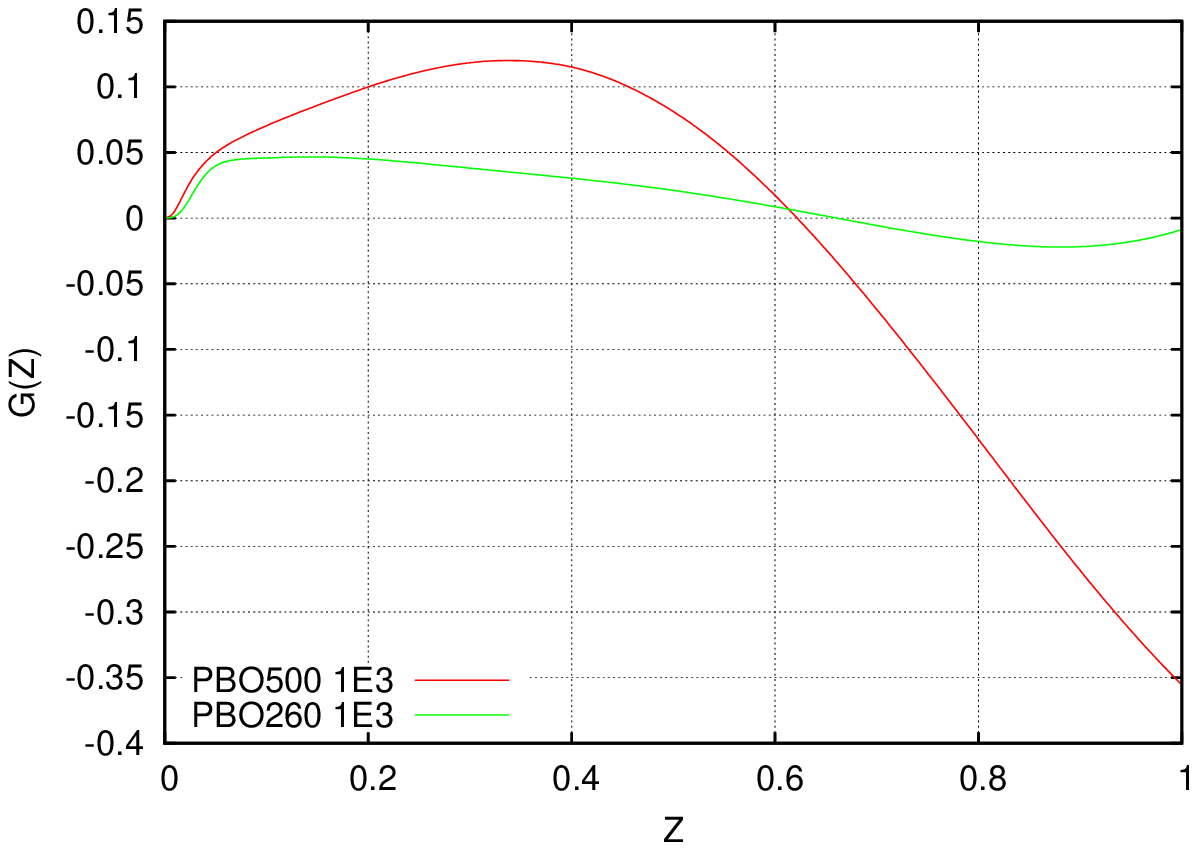}
\caption{Wake function of Couette flow as estimated from the method of \S\ref{wakeestim}.}
\label{couettewakeestim}
\end{figure}

The wake-estimation curves for Couette flow, shown in Figure \ref{couettewakeestim},
exhibit a behaviour markedly different from previous cases, with a much lower positive range ($+0.1$ instead of $+0.7$, effectively zero in the face of foreseeable data errors). The lack of a significant positive overshoot, in the light of the explanation given in \cite{PRL}, can be immediately associated with the value of $g$ being zero. In the corner-defect region we see negative values, consistent with a negative corner-defect function $H(Z)$ which in this case is not offset by any pressure-gradient effect. It is also quite evident that $\Re_\tau=260$ is likely to be too low, and only the difference between the $\Re_\tau=500$ and $\Re_\tau=1000$ profiles can be trusted here.

\begin{figure}
\includegraphics[width=\columnwidth]{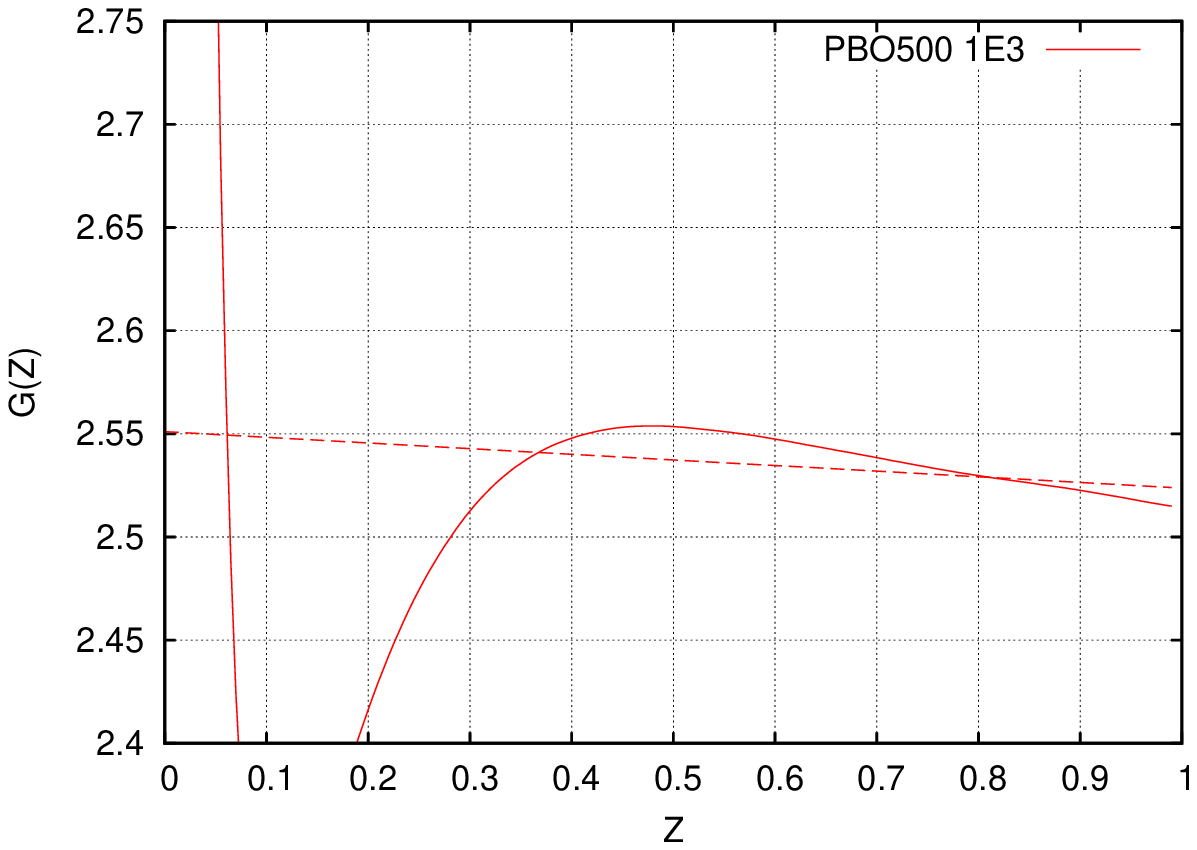}
\caption{$A_0$ and $gA_1$ coefficients of Couette flow as estimated from the method of \S\ref{wakeelim}.}
\label{couetterays}
\end{figure}

The difference between these two profiles in the wake-elimination plot of \S\ref{wakeelim} is shown in Figure \ref{couetterays}. 
This is very satisfactorily consistent with $A_0=1/0.392$ and $gA_1=0$, even if such values could not have been evinced with precision from this plot alone.

\begin{figure}
\includegraphics[width=\columnwidth]{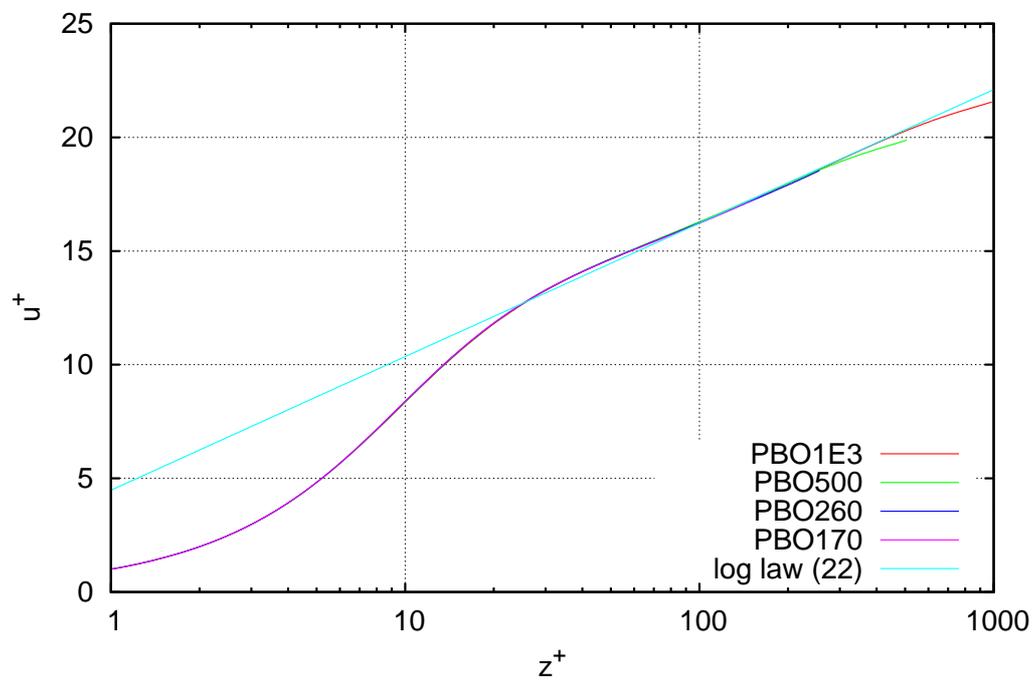}
\caption{Couette velocity profiles compared to the logarithmic law.}
\label{couettef}
\end{figure}

The comparison between Couette velocity profiles (untouched, since their pressure-gradient correction is zero) and the by now standard logarithmic law \eqref{loglaw} is shown in Figure \ref{couettef}.

\begin{figure}
\includegraphics[width=\columnwidth]{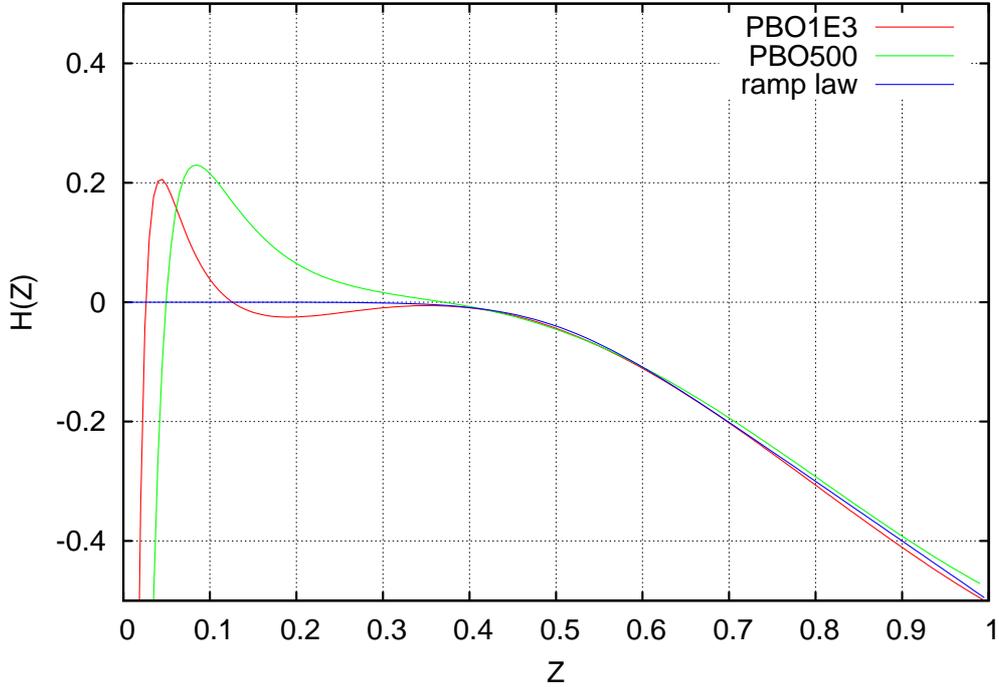}
\caption{Corner-defect function for Couette flow.}
\label{wakefuncc}
\end{figure}

Finally, the corner-defect (actually coincident with the wake) function is shown in Figure \ref{wakefuncc}, as extracted by subtracting the log law from each Couette-flow profile.
As can be seen, the initial kink that appeared in Figure \ref{couettewakeestim} really was an artifact of the method; according to Figure \ref{wakefuncc} the wake function is negative and very nearly monotonic, becomes significant already for $Z\ge 0.4$ (earlier than in the other two geometries), and for practical purposes it is very well approximated by the smoothed-ramp function
\be
H(Z)=\frac{x-0.5}{\exp(-25(x-0.5))-1}
\ee{cwakeH}
Whereas the general shape of the corner-defect function was similar for plane and pipe flow, for Couette flow we see a significant difference. This is unsurprising since here the profile must adapt to a different (odd rather than even) symmetry. It is nonetheless a remarkable coincidence that the maximum negative value of the corner-defect correction ($\sim -0.5$), occurring at the centerline $Z=1$, is similar in all three geometries.

\section{The law of the wall}
\label{wallfunction}
Let us now more closely examine the wall layer, with the purpose of obtaining a description, and possibly some analytical interpolating formula, of the universal law of the wall $f_0(z^+)$ and of its first-order correction  $f_1(z^+)$ in \eqref{unifexp}. Particularly interesting will be how quickly the sum of these functions approaches its overlap behaviour \eqref{newovlap}, so as to know in which range of $z^+$ and with what accuracy the simpler equation \eqref{newovlap} can be relied upon. In doing so we shall zoom in on finer and finer details of the velocity profile, and it should be kept in mind that error of whatever origin is correspondingly amplified and we are venturing in a path where physical effect and numerical (or measurement) error may become indistinguishable, something which is inevitably bound to happen sooner or later when empirical data are being analysed at deeper and deeper resolution.

\begin{figure}
\includegraphics[width=\columnwidth]{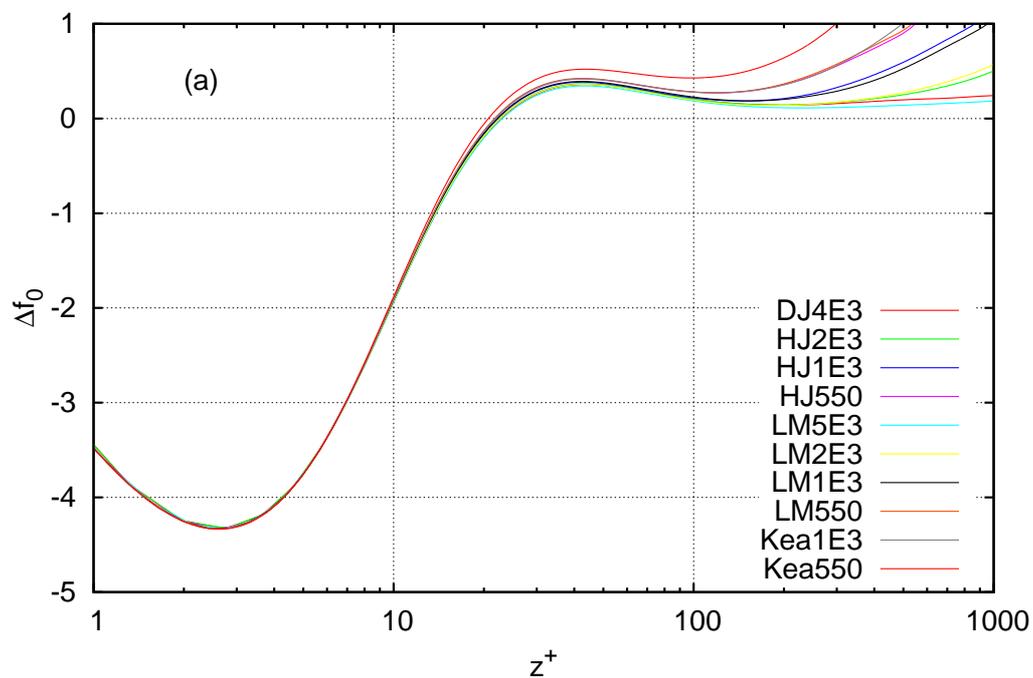}a

\includegraphics[width=\columnwidth]{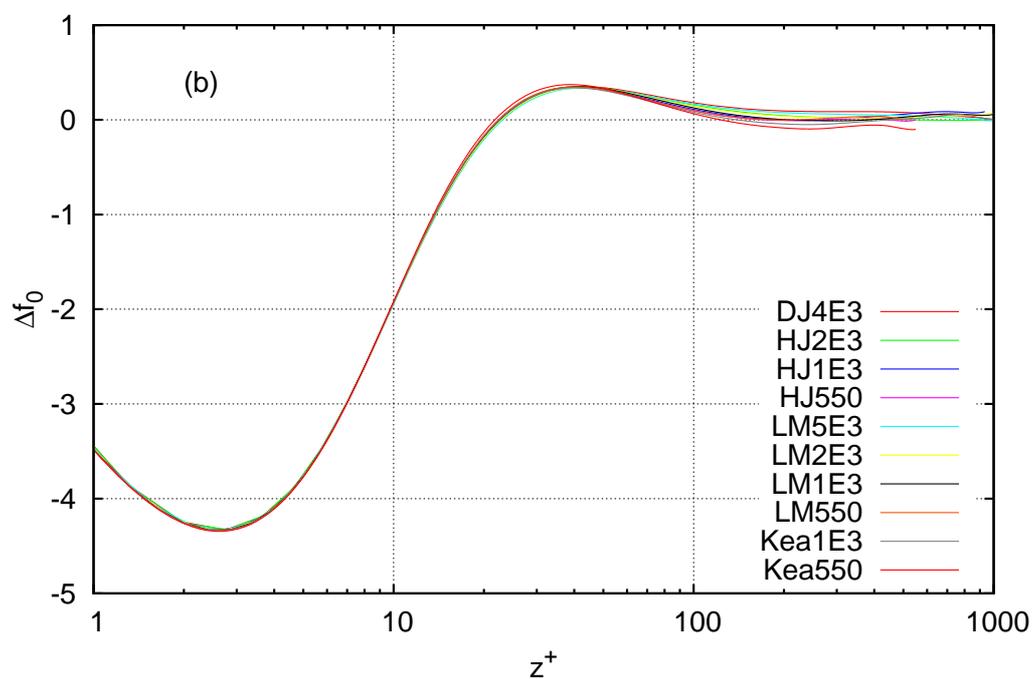}
\caption{Plot (a) of the difference $u^+ - \kappa^{-1}\log(z^+)-B$ and (b) of the difference $u^+ -\kappa^{-1}\log(z^+)-gZ-B$, in wall units.}
\label{wall-log}
\end{figure}

Once the logarithmic law \eqref{loglaw} is taken for granted, the difference between each velocity profile and this log law can be highlighted by rescaling plots similar to Figures \ref{Hwake},\ref{wakefuncp},\ref{wakefuncc} in wall units rather than in outer units.
Such a plot for plane parallel flow is shown in Figure \ref{wall-log}a, where only the logarithmic law is subtracted without any pressure-gradient correction,
and in Figure \ref{wall-log}b where also the pressure-gradient correction is included as
\be
\Delta f_0=u^+ -\kappa^{-1}\log(z^+)-gZ-B .
\ee{deltaf0}

Several noticeable features appear. First of all the agreement between profiles obtained by different authors, at different Reynolds numbers, and in different geometries is very good on the scale of Figure \ref{wall-log}a and even better in Figure \ref{wall-log}b, thus showing that there really is a universal law of the wall. In Figure \ref{wall-log}a we see a linear (as it would appear had we not adopted a logarithmic scale) divergence after $z^+\simeq 100$; as can by now be imagined, this linear divergence is caused by the pressure gradient, and vanishes when the pressure-gradient correction is additionally subtracted in Figure \ref{wall-log}b.
The residual uncertainty in Figure \ref{wall-log}b is of the order of $\pm 0.02$ in its left half and $\pm 0.1$ its right half, a region which will later require some caution.

Secondly, the wall-function difference exhibits a well defined minimum at $z^+\simeq 2.6$ and a well defined maximum at $z^+\simeq  42$, both of which very convincingly scale with wall units. The overshoot of the wall function before approaching the logarithmic law is a remarkable feature, and was highlighted as such before \cite{Monkewitz}, since it eludes the instinctive expectation of a smooth, one-sided approach. %
It also explains the generalized slight overestimation of $\kappa$ (underestimation of $A_0$) in the past, as people were prone to apply a logarithmic law already for $z^+\gtrsim 30$. Nikuradse's original preference for his estimate of $0.417$ over $0.4$ in Figure \ref{Nikupoints} may perhaps also be ascribed to this tendency.
 
One may be tempted to interpret this overshoot as the presence of a ``mesolayer", conjectured in \cite{mesolayer} to possess an intermediate length scale $\sqrt{h\nu/u_\tau}$; however such an interpretation, which may perhaps be left open by a figure such as \ref{wall-log}a, looks unwarranted in the light of Figure \ref{wall-log}b where the position and shape of the maximum clearly stay fixed in wall units with changing Reynolds number.

\subsection{Extrapolation to $\Re_\tau=\infty$}
Whereas it should be apparent that any one of the profiles included in Figure \ref{wall-log}b (or an analytical interpolation of one of those) can already yield a good approximation of the leading-order law of the wall $f_0(z^+)$, or of its increment $\Delta f_0(z^+)$ to be added to \eqref{newovlap}, trying to extract the first-order correction $f_1$ drives us to the next level of magnification.

\begin{figure}
\centering
\includegraphics[height=0.3\textheight]{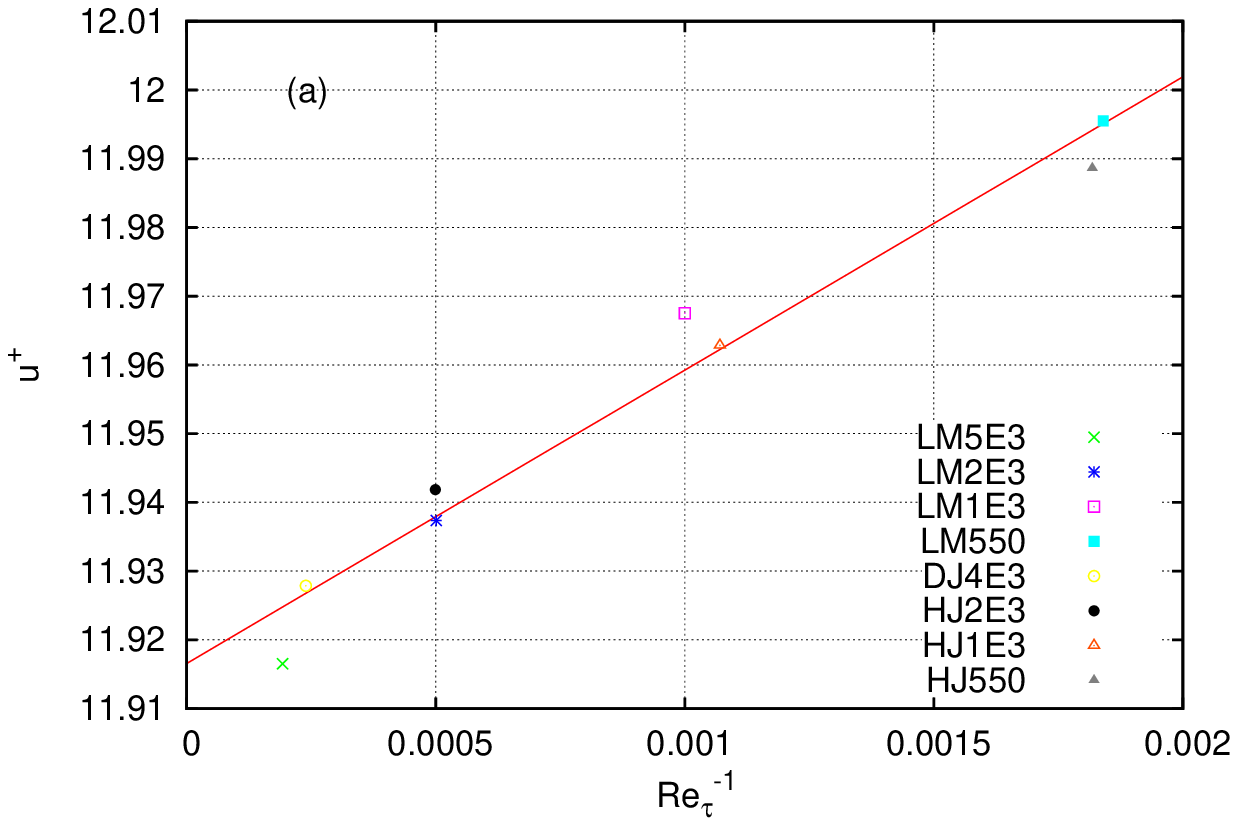}

\includegraphics[height=0.3\textheight]{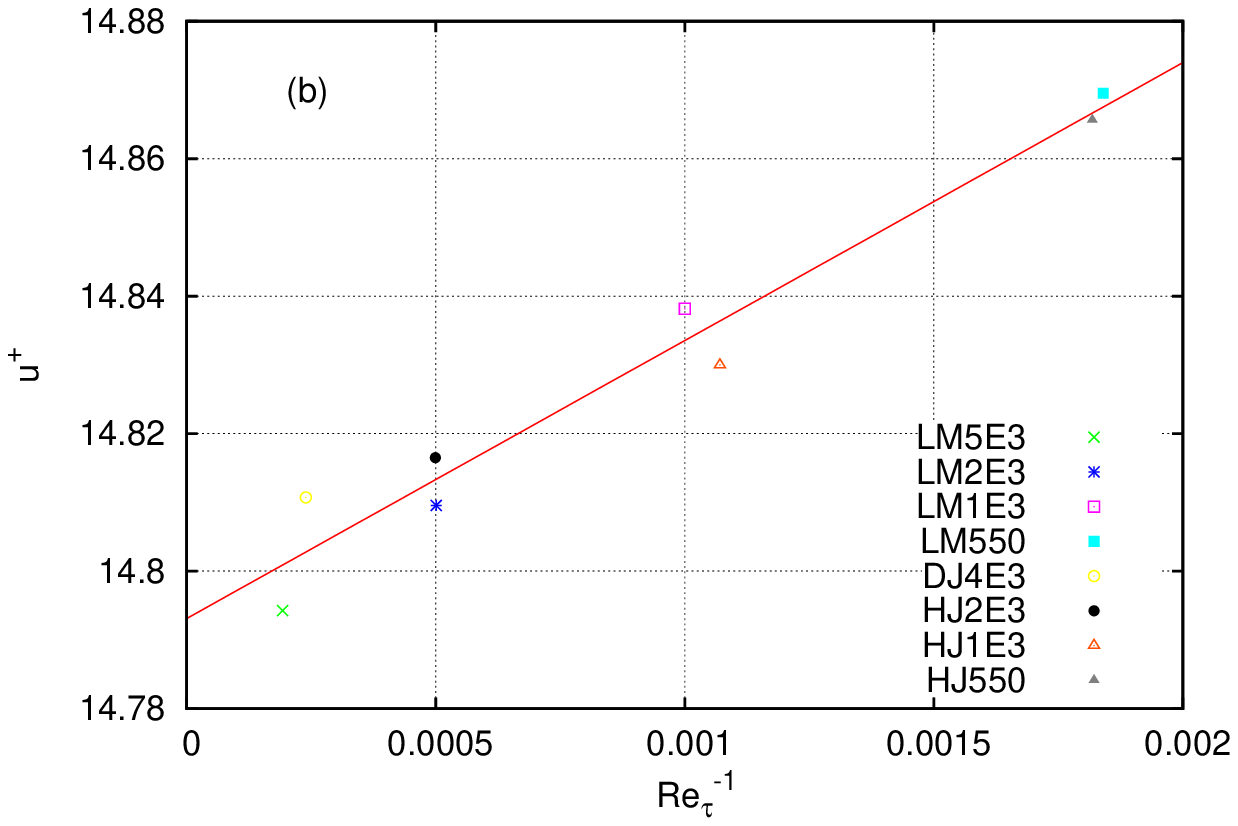}

\includegraphics[height=0.3\textheight]{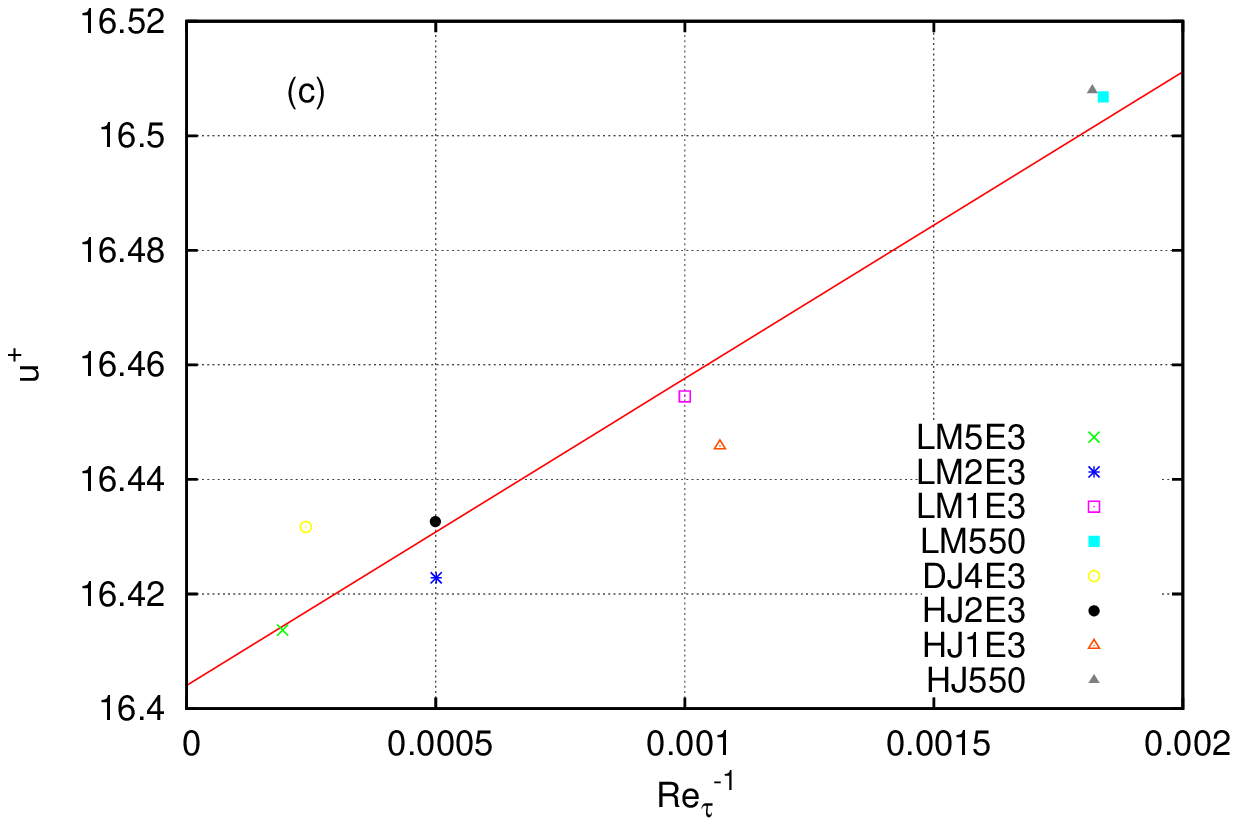}

\caption{$u^+$ as a function of reciprocal Reynolds number for (a) $z^+=20$, (b) $z^+=50$, (c) $z^+=100$.}
\label{islin}
\end{figure}

\begin{figure}
\includegraphics[width=\columnwidth]{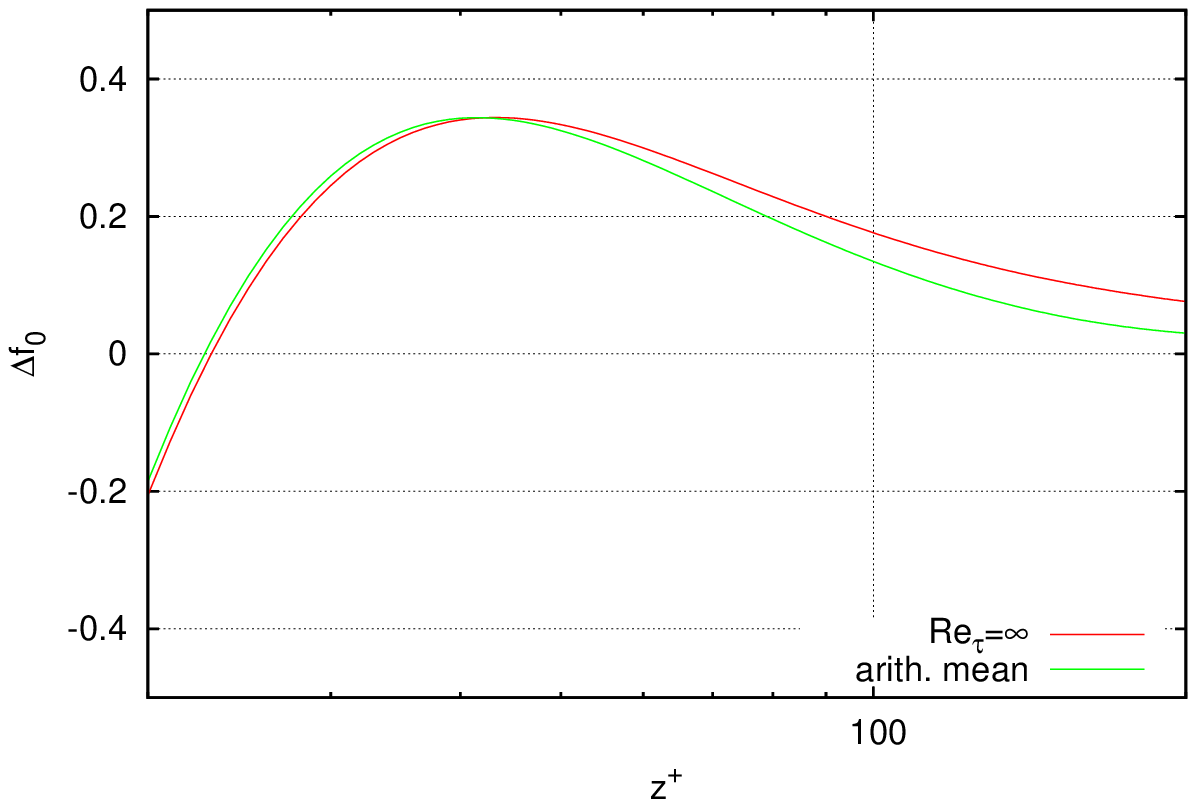}
\caption{Best estimate of the law-of-the-wall increment, in a neighbourhood of its maximum, as obtained either by extrapolating the numerical data to $\Re_\tau=\infty$ or by taking their arithmetic mean.}
\label{extrapf0}
\end{figure}

Truncating \eqref{wallexp} to its first two terms
\be
u^+=f_0(z^+)+\Re_\tau^{-1}f_1(z^+),
\ee{firstterms}
shows that, for constant $z^+$, velocity $u^+$ is expected to be a linear function of $\Re_\tau^{-1}$ (up to some value of $z^+$ where \eqref{firstterms} ceases to be valid). An empirical verification whether this is so is provided by plots such as Figures \ref{islin}a--c, which display velocity as a function of $\Re_\tau^{-1}$ for three different values of $z^+$. As can be seen, the individual points reasonably fit a straight line but not perfectly, and in fact the difference between results of different authors becomes visible on this scale even at positions relatively near to the wall. %
Though Figure \ref{islin} does not afford a precise evaluation of the first-order correction to the wall function, it does allow us to extract its order of magnitude and to state that its absolute value, of $0.1$ in the worst case for $\Re_\tau=500$, is small enough not to be practically very relevant. On the other hand, the linear extrapolation to $1/\Re_\tau=0$ of the data in Figure \ref{islin} looks like the appropriate way, in the absence of additional random errors, to extract the most accurate approximation of $f_0(z^+)$ from the available data files. This extrapolation is shown, in a neighbourhood of its maximum, in Figure \ref{extrapf0} together with a straightforward arithmetic mean of the curves in Figure \ref{wall-log}b. It should be remarked that extrapolation may be dangerous, as it amplifies the non-deterministic part of the error. The arithmetic mean, on the contrary, damps random data errors but introduces a systematic bias. 

As can be seen, the difference between the two estimates crosses zero near their maximum, located at $z^+\simeq 42$, and grows larger afterwards where, however, the spread among data in Figure \ref{wall-log}b was also larger. Caution suggests as a compromise to adopt the extrapolated curve up to the crossing point and the averaged curve thereafter, and this will be our present best approximation of the law-of-the-wall increment $\Delta f_0$ defined by \eqref{deltaf0}.

\subsection{Interpolation}
In lack of a closed physical model of mean turbulent flow, obvious practical convenience has always made analytical interpolations of the law of the wall desirable. A very early such interpolation was devised by Spalding \cite{Spalding}, in the form of a monotonic function satisfying the end conditions $f_0(0)=0$, $f'_0(0)=1$, $f''_0(0)=0$, $f'''_0(0)=0$, and $f(z^+)\rightarrow \kappa^{-1}\log(z^+)+B$ for $z^+\rightarrow\infty$. He obtained this result by first inverting the logarithmic law \eqref{clasovlap} as $z^+=\e^{\kappa (u^+ - B)}$, and then subtracting as many terms of the Taylor series of this exponential as required to zero the appropriate number of derivatives:
\be 
z^+=u^+ +\e^{-\kappa B}\left({\e^{\kappa u^+}} - \kappa u^+ -\frac{\kappa^2 {u^+}^2}{2}-\frac{\kappa^3 {u^+}^3}{6}\right)
\ee{Spaldingeq}

Spalding's formula is elegant but implicit, as \eqref{Spaldingeq} needs to be inverted to give $u^+$ as a function of $z^+$. As will be seen below, it is also the least accurate.

An explicit formula was provided by Musker \cite{Musker}, who started with a rational (ratio of two polynomials) expression of Reynolds stress, designed to satisfy appropriate asymptotic conditions at $z^+=0$ and $z^+\rightarrow\infty$, and then analytically integrated it to extract the velocity profile as the formula
\begin{multline}
u^+=5.424\arctan\frac{2z^+-8.15}{16.67}%
+\log_{10}\frac{(z^+ +10.6)^{9.6}}{({z^+}^2-8.15z^+ +86)^2}-3.52 .
\label{Muskereq}
\end{multline}

Neither Spalding nor Musker used other data than behaviour near zero and infinity to constrain their interpolation. More recently, Monkewitz \etal\ \cite{Monkewitz} adopted a rational Padé representation of the Reynolds stress formally similar to Musker's, but using a larger number of free coefficients, and fitted it to a set of experimental profiles measured in a zero-pressure-gradient boundary layer. (Like Musker's, this was part of a more extensive effort aimed at describing both the wall and wake regions of a turbulent boundary layer. Nonetheless, the part of their analysis concerning the wall layer also applies to parallel flow insofar as the wall layer is universal.) Monkewitz \etal\ \cite{Monkewitz} split the Reynolds stress in the sum of two contributions, using a Padé function P$_{23}$, ratio of a polynomial of degree $2$ and one of degree $3$, for one and a Padé function P$_{25}$ for the other, and determined the coefficients of these two rational functions from a combination of boundary conditions and experimental data points. An explicit expression of the velocity profile requires analytically integrating the Reynolds stress, and takes the form of a sum of several logarithms and arctangents of polynomials (Eq. (6) of \cite{Monkewitz}).

In order to develop an interpolation of the data in Figure \ref{extrapf0}, we preliminarily observe that the integral of a rational function will always have an asymptotic behaviour at infinity given by the integral of its Laurent power series, namely one that can be expressed as a single logarithm of a first-degree polynomial plus another rational function. Therefore, in order to set up a generic formula with free coefficients apt to fit empirical data, we would get a more easily manageable expression if, instead of integrating a rational function like in the above examples, we wrote $f_0(z^+)$ as the sum of a logarithm and a Padè approximant and fitted this sum to the data.

As an additional consideration, either way of using rational functions still restricts their asymptotic behaviour to an algebraic form, which is another name for a Laurent series. This is a very specific kind of asymptotic behaviour, involving analyticity in the complex plane, and among the differential equations of mathematical physics it is a prerogative of the Laplace equation in open space. In many other physical phenomena, such as those governed by constant-coefficient differential equations (including the Laplace equation between walls) an exponential approach to their behaviour at infinity is more common. Particularly in the matched asymptotic expansion of boundary layers \cite{vanDyke}, to which the asymptotic extensions of Millikan's theory are inspired, approach of the inner-layer solution to its own asymptotic behavior is exponential. 

Whereas a Padé interpolation over an infinite interval will eventually converge given enough free coefficients, it will do so slowly if it does not match the appropriate asymptotic behavior. In order to empirically decide whether the data in Figure \ref{extrapf0}
\begin{figure}
\includegraphics[width=\columnwidth]{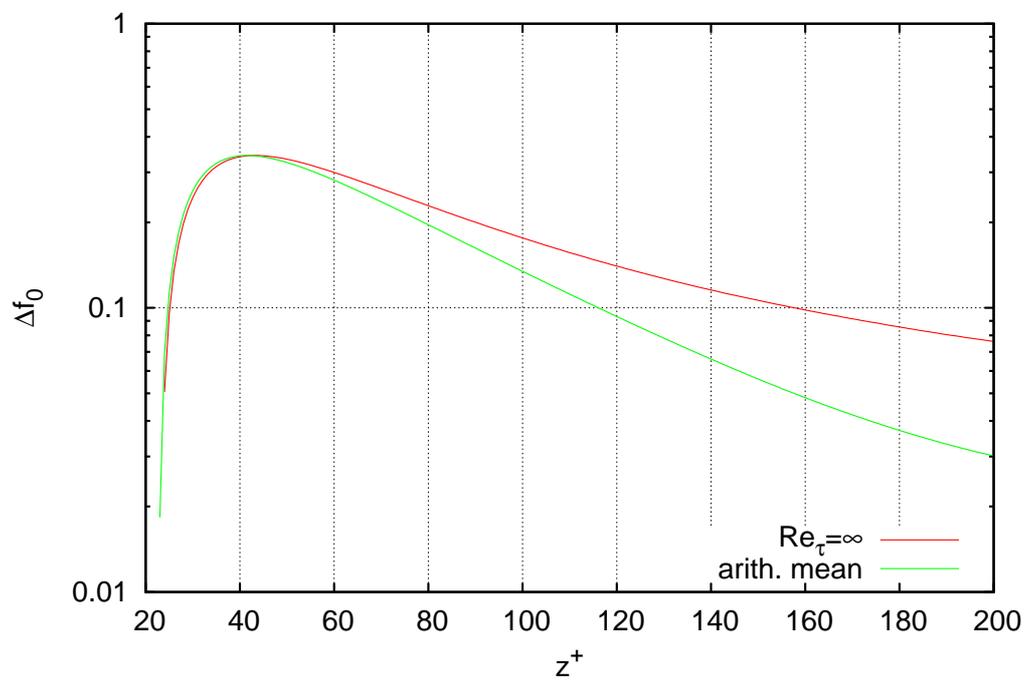}
\caption{Extrapolated and mean law-of-the-wall increment in logarithmic vertical scale.}
\label{extrapf0log}
\end{figure}
look more like exponential rather than algebraic, a simple device is to replot them on a logarithmic vertical scale, as done in Figure \ref{extrapf0log}.
The appearance of about a decade of nearly constant slope provides an indication that the asymptotic decay of $\Delta f_0$ is more likely to be exponential than algebraic. (Or at least, aside from any theoretical speculations, an exponential interpolant is more likely to succeed in interpolating this curve.) Balancing this consideration with the undoubtful practical convenience of rational functions, our final choice is a P$_{22}$ Padé approximant multiplied by an exponential, summed to the logarithm of a first-degree polynomial:
\begin{multline}
f_0(z^+)=\frac{a_0+a_1{z^+}+a_2{z^+}^2}{1+b_1{z^+}+b_2{z^+}^2}\,e^{-c{z^+}}%
+\kappa^{-1}\log(z^++d)+B
\label{interpolant}
\end{multline}
Imposing the initial conditions $f_0(0)=0$ and $f'_0(0)=1$, together with the asymptotic values $\kappa=0.392$ and $B=4.48$, and performing a least-square fit of the remaining 5 coefficients to the data of Figure \ref{extrapf0}, yields the values listed in Table \ref{pade}.
\begin{table}
\caption{Coefficients of the exponential-Padé approximant of the law of the wall.}
\label{pade}
\[
\begin{array}{rcl}
a_0&=              & -7.374\\
a_1&=              & -0.4930\\
a_2&=              & 0.02450\\
b_1&=             & 0.05736\\
b_2&=              & 0.01101\\
c&=               & 0.03385\\
d&=               & 3.109\\
\end{array}
\]
\end{table}

\begin{figure}
\includegraphics[width=\columnwidth]{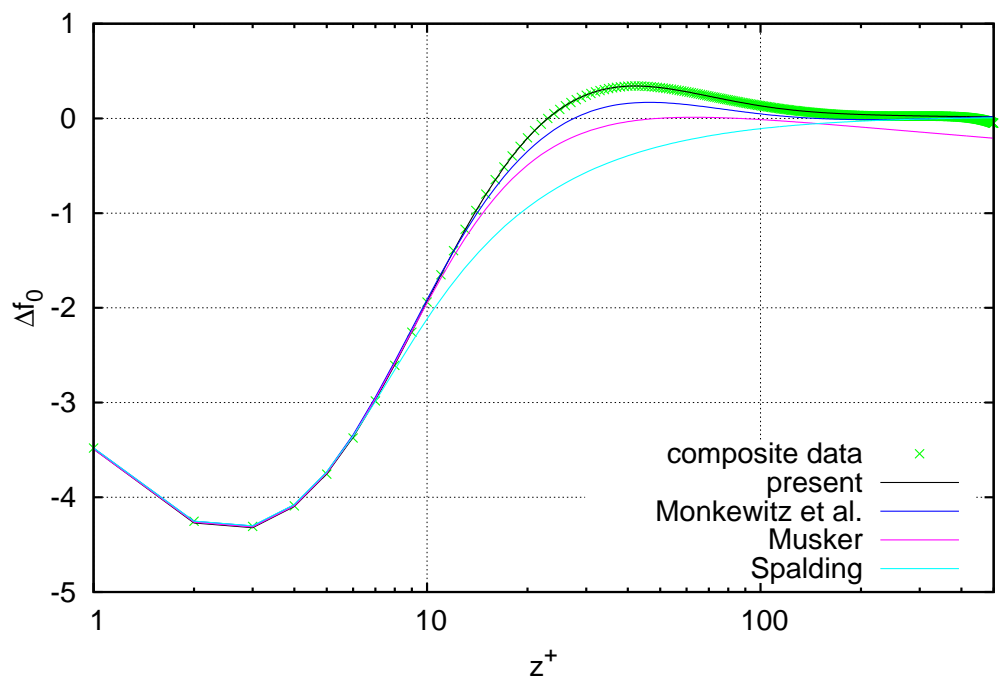}
\caption{Different approximations of the law of the wall}
\label{varinterp}
\end{figure}
\begin{figure}
\includegraphics[width=\columnwidth]{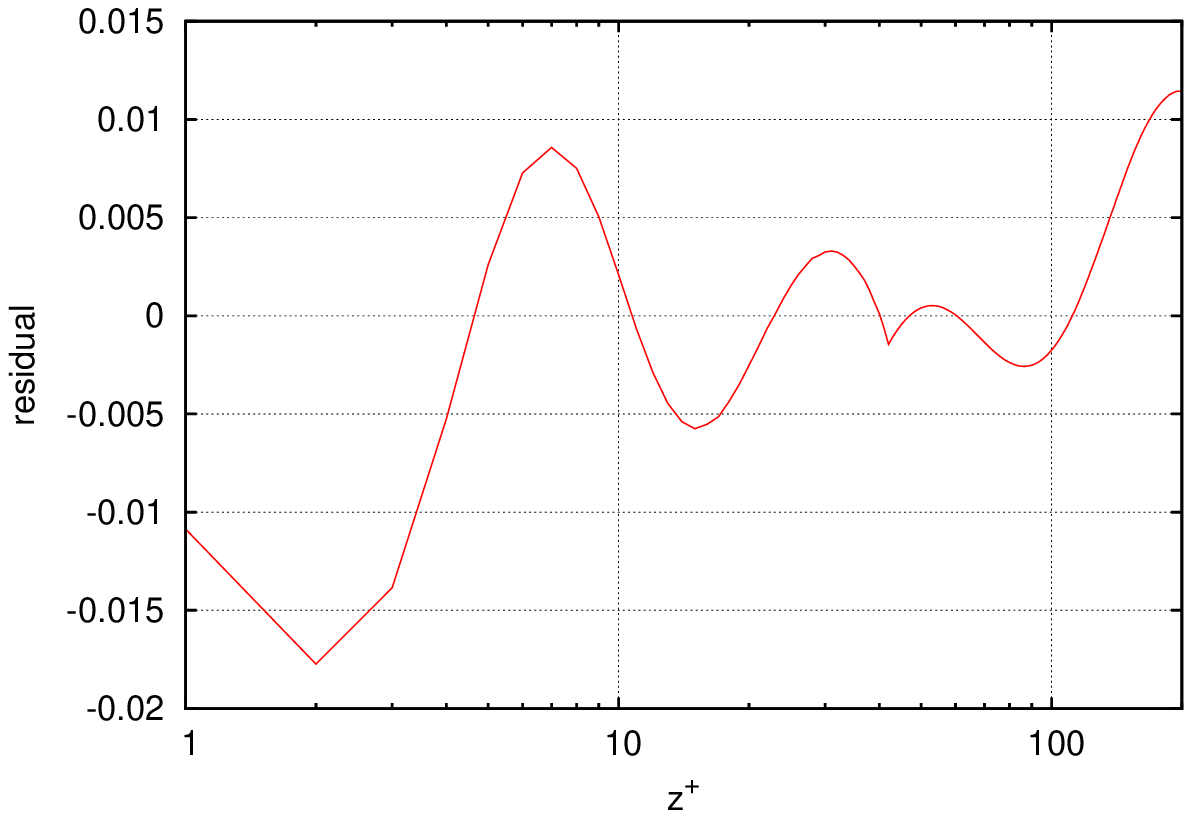}
\caption{Residual difference between interpolant formula \eqref{interpolant} and the composite data of Figure \ref{extrapf0}.}
\label{interpresid}
\end{figure}

A plot of \eqref{interpolant} is shown in Figure \ref{varinterp}, compared to the composite extrapolated and averaged data of Figure \ref{extrapf0} and to the past approximations of \cite{Monkewitz,Musker,Spalding}, each represented as a wall-function increment $\Delta f_0$ obtained by subtracting one and the same log law \eqref{loglaw}. %
 The error of the interpolant \eqref{interpolant} with coefficients from Table \ref{pade}, undistinguishable from zero in Figure \ref{varinterp}, is further displayed in Figure \ref{interpresid} and is of the order of $\pm 0.01$ in wall units.

\section{Roundup and conclusion}
After going through the discussion in the previous sections, one sees that our present best approximation to the mean velocity profile in different geometries of parallel flow still fits within Coles' shrewd formula
\be
u^+=f_0(z^+)+G(Z)
\ee{finColes}
where, however, there are now explicit interpolations available of the universal wall function $f_0(z^+)$ and of each specific wake function $G(Z)$. $f_0(z^+)$ is for all practical purposes well represented by its analytical interpolant \eqref{interpolant}, and $G(Z)$ is given by
\be
G(Z)=gZ+H(Z),
\ee{finwake}
$gZ$ being the effect of the pressure gradient\footnote{More precisely, as shown in \cite{PRL}, the effect of the pressure gradient must be written as $A_1 g Z$ where $A_1$ is a universal constant, derived from the same kind of dimensional argument as von Kármán's constant is, the best approximation of which is presently $A_1=1$.},
with the geometry factor $g$ equal to $2$ for pipe flow, $1$ for plane duct flow, and $0$ for turbulent Couette flow, and $H(Z)$ being the corner-defect function approximated by, respectively, \eqref{pwakeH}, \eqref{Hintrp}, and \eqref{cwakeH} for each geometry.

An even better representation must in principle be the higher-order uniformly-valid formula
\be
u^+=f_0(z^+)+G(Z)+\Re_\tau^{-1}f_1(z^+),
\ee{higherColes}
but an estimate of $f_1(z^+)$ sits on the edge of what can be
gotten within the accuracy of present simulations and
experiments. This additional contribution to $u^+$ is, at any rate, of less than $0.1$ absolute value for $\Re_\tau\ge 500$. %

\begin{figure}
\includegraphics[width=\columnwidth]{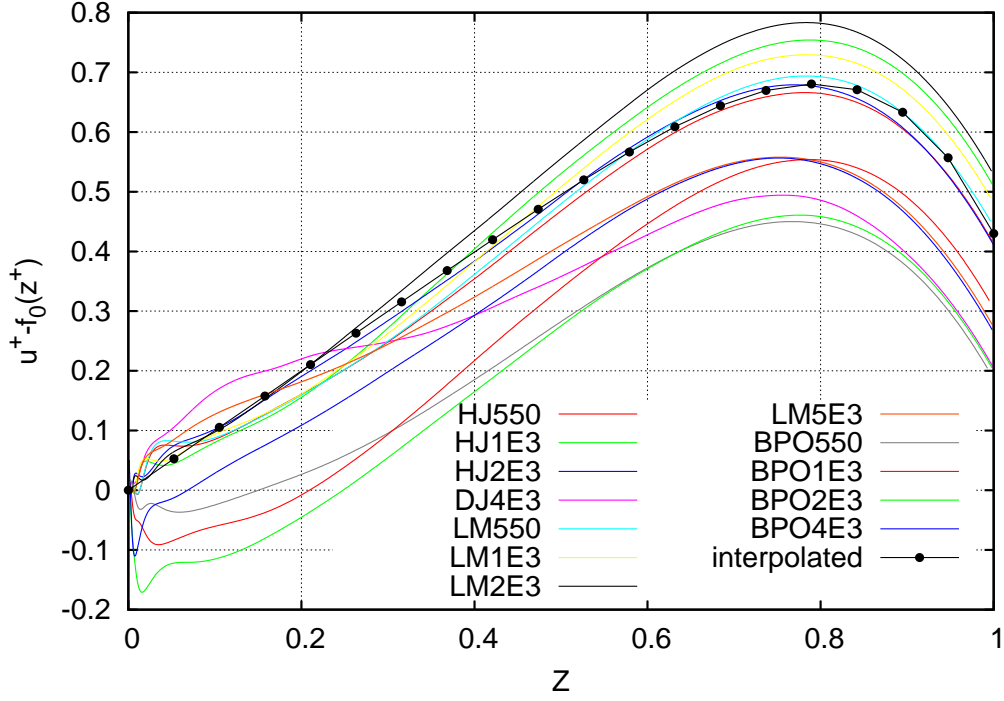}
\caption{A-posteriori reconstructed wake function for plane-duct flow.}
\label{postwakeduct}
\end{figure}

\begin{figure}
\includegraphics[width=\columnwidth]{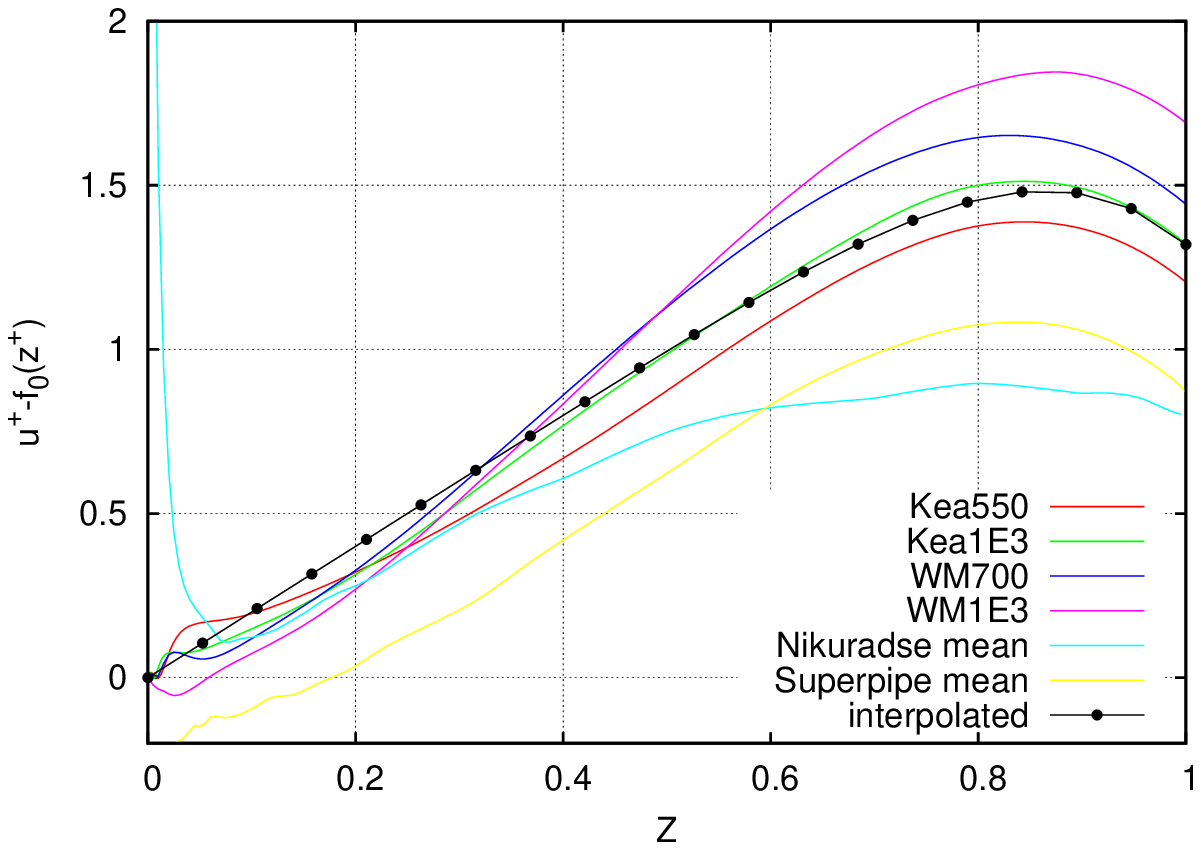}
\caption{A-posteriori reconstructed wake function for pipe flow.}
\label{postwakepipe}
\end{figure}

\begin{figure}
\includegraphics[width=\columnwidth]{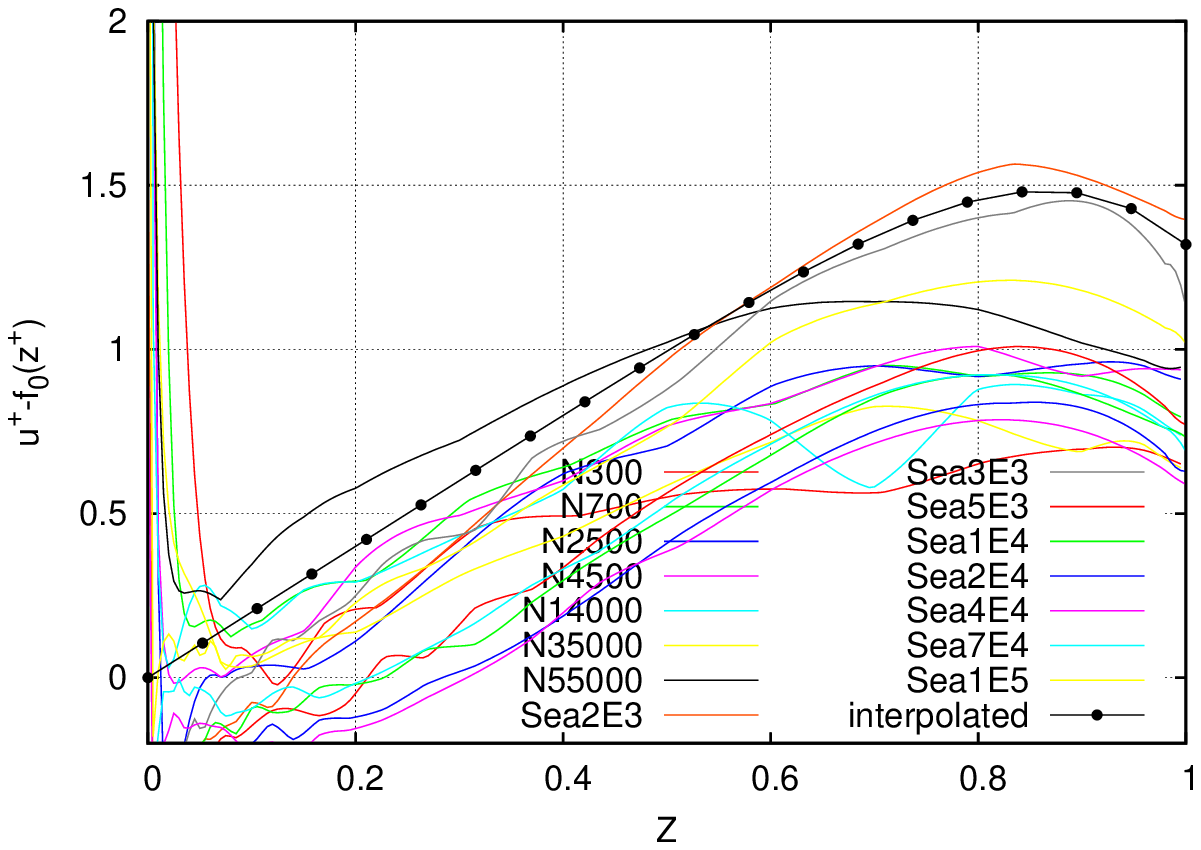}
\caption{A-posteriori reconstructed wake function for pipe flow experiments.}
\label{postwakeexppipe}
\end{figure}

\begin{figure}
\includegraphics[width=\columnwidth]{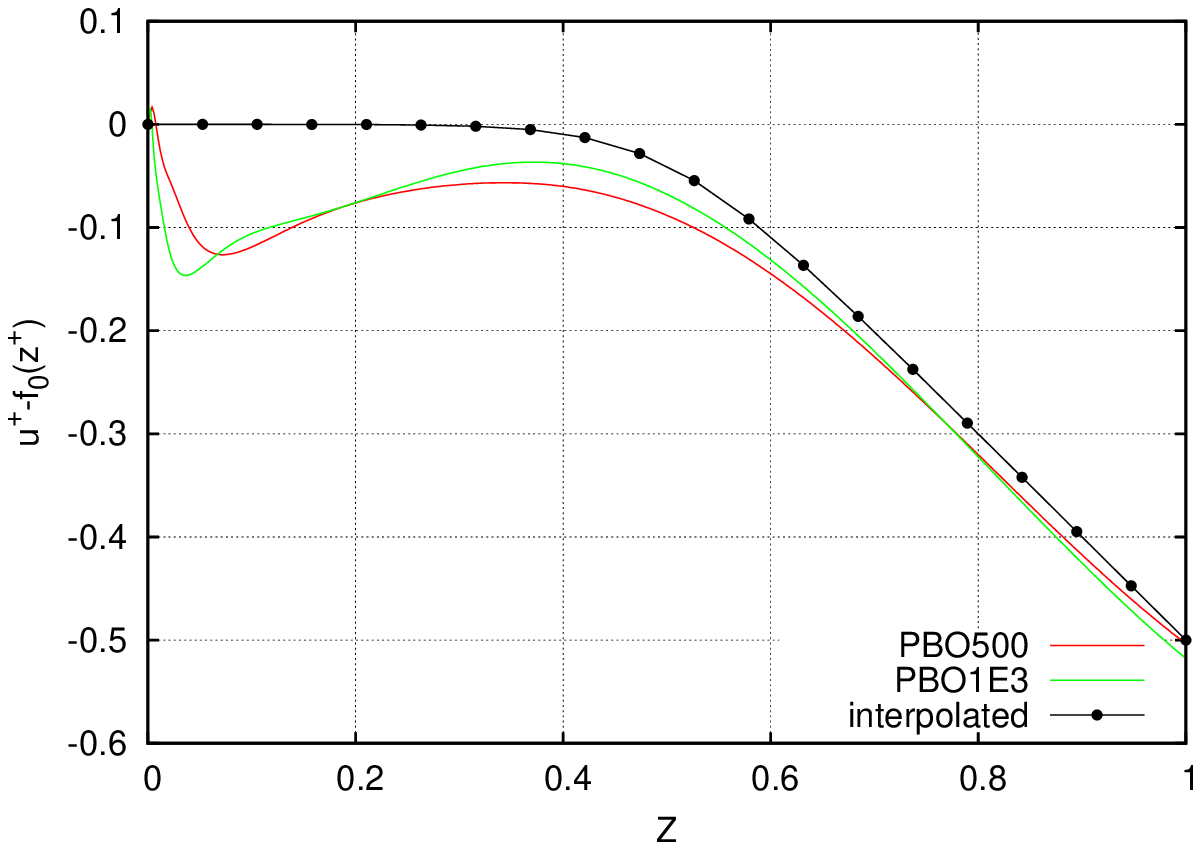}
\caption{A-posteriori reconstructed wake function for Couette flow.}
\label{postwakecouette}
\end{figure}

In order to extract the wall function $f_0(z^+)$ from empirical data we have, in practice, first extracted an approximation of the wake function $G(Z)$ from the comparison of more than one velocity profile at different Reynolds numbers, and then subtracted $G(Z)$ from one of those. The technique to do this was explained in \S\ref{wakeestim}, consisting in the solution of a functional equation involving the difference of two velocity profiles taken at the same wall-units coordinate $z^+$. The method is computationally easy, and provides a visual image of the wake function that does not require any assumption other than the validity of \eqref{finColes}; it also shows that the extracted wall function actually becomes logarithmic for $z^+\gtrsim 200$, thus the range of Reynolds numbers for which numerical simulations are available is at all sufficient in order to try and estimate the value of von Kármán's constant. 

Unfortunately, this estimate still contains too much error (data uncertainty) to provide a sufficiently precise value of $\kappa$. Therefore in \S\ref{wakeelim} we proposed a different method that, by taking differences of velocity profiles at the same outer coordinate $Z$ rather than at constant $z^+$, mathematically eliminates the wake function even if doing so hides its shape. The outcome of this method, Figure \ref{rays0} in the case of plane-duct flow, is expected to be a set of superposed horizontal straight lines pointing at the true value of $\kappa$, but much to our surprise it turned out to be close to a pencil of inclined straight lines radiating from a single point located at $A_0=1/0.392$. The inclination of such lines suggests that the different profiles have somewhat different wake functions, and in the light of the theory \cite{PRL} which associates the slope of the wake function with the pressure gradient, it is as though the numerical error mimicked somewhat different pressure gradients. (Alternately, this might be a true physical effect that makes \eqref{finColes} invalid, but the difference being of the same order of magnitude as the difference between results of different authors makes us lean towards numerical error as the source.)

Pending a verification with more accurate data in the future, we settled on $\kappa=0.392$ and $A_1=1$ as our present best estimates. We then proceeded to examining various datasets of plane-duct, circular-pipe and turbulent-Couette flow in the light of this logarithmic law and extracted both the wake function of each geometry and a common law of the wall, for which the compact analytical interpolation \eqref{interpolant} was provided.

Having decided on what $f_0(z^+)$ should look like, we can now go full turn and test the validity of our previous assumptions, and at the same time the accuracy of the data, by reconstructing the wake function of each velocity profile as $G(Z)=u^+-f_0(z^+)$. This is done in Figures \ref{postwakeduct}-\ref{postwakecouette}, where the theoretical wake function, as deduced in the corresponding section of this article, is also superposed. As can be seen there is a considerable residual spread of the order of $\pm 0.2$ for a plane duct and even larger for pipe flow (we cannot comment about spread for Couette flow since a single data set is available in this case), but the fact that the predicted wake function, using one and the same wall function for all three geometries, fits within the data cloud in all cases is a nontrivial confirmation.

The open question is, of course, is the residual spread just caused by error (and therefore will decrease with future improvements in numerical and experimental measurement techniques), or does it accomodate some yet undiscovered physical feature? Closer inspection reveals a possible pattern in these figures, behind the evident random scatter, of general decrease of $G$ with Reynolds number. This pattern might point at a slightly revised estimate of $\kappa$, or at a different shape of the wake function (perhaps a quadratic at small $Z$, as Coles \cite{Coles} seemed to expect, rather than a linear function?), or at the presence of higher-order corrections of the form \eqref{unifexp}. But it might also point at some systematic error in the numerical techniques (non-uniform discretization or size of the periodic box for instance, just to mention two whose influence changes with Reynolds number) or in the experimental techniques. It is in the human nature to look for patterns, and strong is the temptation to speculate further, but such a chase for finer and finer details can never end and easily go astray. We prefer to content ourselves with the present level of investigation and wait for future data improvements, with the only final remark that the techniques and plots described in this paper can also become a useful tool to analyse the data themselves and their accuracy during their retrieval.

\end{document}